\documentclass[submission, Phys]{SciPost}

\usepackage{color,soul}
\hypersetup{
    colorlinks,
    linkcolor={red!50!black},
    citecolor={blue!50!black},
    urlcolor={blue!80!black}
}

\usepackage[bitstream-charter]{mathdesign}
\DeclareSymbolFont{usualmathcal}{OMS}{cmsy}{m}{n}
\DeclareSymbolFontAlphabet{\mathcal}{usualmathcal}

\usepackage{braket}

\newcommand{\abs}[1]{\left| #1 \right|}

\definecolor{indiagreen}{rgb}{0.07, 0.53, 0.03}

\begin{document}

\begin{center}
    {\Large \textbf{
    Crossover from attractive to repulsive induced interactions and\\ bound states of two distinguishable Bose polarons\\}}
\end{center}

% Separate subsequent authors by a comma, omit comma and use "and" for the last author.
% Mark the corresponding author with a superscript star.
\begin{center}
    Friethjof Theel\textsuperscript{1},
    Simeon I. Mistakidis\textsuperscript{2,3} and
    Peter Schmelcher\textsuperscript{1,4}
\end{center}

\begin{center}
    {\bf 1} Center for Optical Quantum Technologies, University of Hamburg, Department of Physics, Luruper Chaussee 149, D-22761, Hamburg, Germany
    \\
    {\bf 2} ITAMP, Center for Astrophysics $|$ Harvard $\&$ Smithsonian, Cambridge, MA 02138 USA
    \\
    {\bf 3} Department of Physics, Harvard University, Cambridge, Massachusetts 02138, USA
    \\
    {\bf 4} The Hamburg Centre for Ultrafast Imaging, University of Hamburg, Luruper Chaussee 149, D-22761, Hamburg, Germany
\end{center}

\begin{center}
    \today
\end{center}

\section*{Abstract}
{\bf
We study the impact of induced correlations and quasiparticle properties by immersing two distinguishable impurities in a harmonically trapped bosonic medium. It is found that when the impurities couple both either repulsively or attractively to their host, the latter mediates a two-body correlated behavior between them. In the reverse case, namely the impurities interact oppositely with the host, they feature anti-bunching. Monitoring the impurities relative distance and constructing an effective two-body model to be compared with the full many-body calculations, we are able to associate the induced (anti-) correlated behavior of the impurities with the presence of attractive (repulsive) induced interactions. Furthermore, we capture the formation of a bipolaron and trimer state in the strongly attractive regime. The trimer refers to the correlated behavior of two impurities and a representative atom of the bosonic medium and it is characterized by an ellipsoidal shape of the three-body correlation function. Our results open the way for controlling polaron induced correlations and creating relevant bound states.
}

\vspace{10pt}
\noindent\rule{\textwidth}{1pt}
\tableofcontents\thispagestyle{fancy}
\noindent\rule{\textwidth}{1pt}
\vspace{10pt}

\section{Introduction}
\label{sec:introduction}

Impurities embedded in a many-body medium, e.g. a Bose-Einstein condensate (BEC), are dressed by its excitations and generate quasiparticles \cite{massignan2014, schmidt2018a}. In the case of a structureless host these refer to polarons~\cite{landau1933}, while for instance, utilizing a magnetic environment or in the presence of a cavity, magnetic polarons \cite{ashida2018, mistakidis2022,rammelmuller2023magnetic} and polaritons \cite{grusdt2016, kockum2019a} are formed respectively. 
Polarons, which we will investigate herein, have been widely studied in cold-atom settings owing to the enormous flexibility, e.g., in terms of controlling the spatial dimension \cite{grimm2000, bloch2008, catani2012}, the interparticle interactions~\cite{olshanii1998, kohler2006, chin2010}, as well as the trapping geometry and the number of species \cite{haas2007, taglieber2008, papp2008, wu2011, barker2020} and atoms \cite{serwane2011, lester2018}. 
Depending on the statistics of the medium both Bose \cite{catani2012, fukuhara2013, hu2016, jorgensen2016, meinert2017, skou2021} and Fermi \cite{schirotzek2009, kohstall2012, massignan2014} polarons have been experimentally realized, while theoretically fundamental properties of these type of quasiparticles including effective mass \cite{penaardila2015, grusdt2017a, jager2020}, residue \cite{massignan2014, schmidt2018a}, and bound state formation \cite{massignan2014, schmidt2018a, camacho-guardian2018} emerging in two-component systems have been discussed.
Interestingly, by immersing at least two impurities into a quantum gas the latter mediates interactions between the former~\cite{klein2005, brauneis2021,petkovic2022,mistakidis2022cold}, a phenomenon that has been interpreted in terms of a Casimir-type interaction describing the induced interaction between two objects in a fluctuating medium \cite{schecter2014, reichert2019, reichert2019a}.
In particular, induced interactions between two impurities are solely attractive as long as they couple in the same way (i.e. in terms of sign and strength) to the fluctuating medium~\cite{recati2005, pavlov2018, dehkharghani2018, camacho-guardian2018a, pasek2019, reichert2019, reichert2019a, will2021, astrakharchik2023}.
The magnitude of this induced attraction, in general, increases for larger impurity-medium coupling strength and specifically for sufficiently strong attractive ones the impurities assemble in a bound state that can be a bipolaron~\cite{casteels2013, camacho-guardian2018,camacho-guardian2018a, naidon2018, will2021, jager2022} or a trimeron \cite{nishida2015}. Notice that besides the above-discussed studies in a homogeneous BEC environment, the attractive nature of induced interactions has been unveiled also for a harmonically confined~\cite{dehkharghani2018,theel2022,mistakidis2020a} or a lattice trapped~\cite{yordanov2023} medium.
Moreover, in the context of open quantum systems where, e.g., two non-interacting particles are coupled to a heat bath, a mediated induced entanglement between the particles has been predicted and its interplay with the inherent decoherence effects has been analyzed for instance in terms of the interatomic distance and temperature~\cite{braun2002, benatti2003, horhammer2008, duarte2009, zell2009, shiokawa2009, benatti2010, fleming2012, charalambous2019}.

Interestingly, it was predicted~\cite{schecter2014, reichert2019a} that there is also the possibility of mediating repulsive impurity-impurity interactions when two impurities are coupled with different signs to a bosonic bath. In this sense, the underlying experimentally relevant three-component system~\cite{taglieber2008, wu2011} allows to unravel additional polaronic properties as it has been also argued by immersing impurities into a two-component pseudospinor mixture~\cite{compagno2017, ashida2018, charalambous2020, keiler2021, mistakidis2022, stefanini2023,rammelmuller2023magnetic} in order to create, for instance, spin-wave excitations and magnetic polarons~\cite{ashida2018,mistakidis2022}, impurities diffusive response~\cite{charalambous2020} or to facilitate the detection of the dressing cloud via interferometry \cite{ashida2018}. 
However, quasiparticle formation in three-component systems is largely unexplored, besides the few above-mentioned recent studies. An interesting direction is to exploit the tunability of such mixtures, e.g. in terms of different intercomponent couplings, for devising the ground state quasiparticle properties such as the impurities effective mass and induced interactions. Here, it is important to understand the interplay of the latter properties and the underlying impurities' correlations. Also, the formation of relevant bound states either solely among the impurities (bipolarons) or between the impurities and the host atoms (trimers) remains elusive. 
To address these questions, we consider two distinguishable and non-interacting impurities that are embedded into an one-dimensional bosonic gas. The impurities' couplings with the host are individually tuned spanning the regime from attractive to repulsive interactions. Here, the effective interactions between the impurities can be only mediated in the presence of impurity-medium entanglement and bound states require the involvement of strong correlations. As such, to account for the relevant inter and intra-component correlations we employ the variational multilayer multiconfiguration time-dependent Hartree method for atomic mixtures (ML-MCTDHX) approach~\cite{kronke2013, cao2013, cao2017} which is well established for investigating impurity physics~\cite{mistakidis2022cold}. 

Inspecting the spatial two-body correlations between the two impurities we reveal that, in general, they are correlated (anti-correlated) when the two impurity-medium coupling strengths posses the same (opposite) sign. To shed more light on the impact of induced impurities' correlations we carefully monitor their relative distance~\cite{mistakidis2020b}, excluding all mean-field type contributions, for varying coupling strengths. A central result of our work is that the impurities' correlated (anti-correlated) behavior is related to a decrease (increase) of their relative distance, thus, indicating the presence of an induced attraction (repulsion) between them. This observation is additionally confirmed by constructing an effective two-body model in the weak impurity-medium coupling regime inspired from the case of indistinguishable impurities \cite{mistakidis2020a,mistakidis2022cold}. It specifically allows to assign the impurities' induced interaction strength and sign but also other quasiparticle related properties such as their effective mass and trap frequency.

For strong impurity-medium attractions, we identify the formation of a bipolaron state involving the two distinguishable impurities. This bound quasi-particle state is characterized by the so-called bipolaron energy \cite{camacho-guardian2018}, and the size of the impurities' dimer state featuring an exponential decrease for larger attractions. Proceeding a step further, we find that for such strong attractive impurity-medium interactions the three-body correlation function features an ellipsoidal shape indicating bunching and revealing the creation of a trimer state among the two impurities and a corresponding bath atom. To further testify the existence of this trimer state we employ the Jacobi relative distances of the three distinguishable atoms~\cite{greene2017} showing an exponentially decreasing trend for increasing impurity-medium attractions. 

This work is organized as follows. In section~\ref{sec:setup}, the three-component setup under consideration is introduced and in Section~\ref{sec:mlx-method} we explain the variational method used to obtain the ground state properties of the many-body system. Section~\ref{sec:1bd} elaborates on the possible ground state density configurations upon varying the impurity-medium couplings. The emergence of induced impurity-impurity correlation patterns is explicated in Section~\ref{sec:corr}. The interrelation of the aforementioned induced correlations with the induced attractive and repulsive impurity interactions is provided in Section~\ref{sec:induced_interactions} through monitoring their relative distance and constructing an effective two-body model. Delving into the strongly attractive impurity-medium interaction regime, we demonstrate the formation of a bipolaron state among the two distinguishable impurities in Section~\ref{sec:bip} and the generation of a trimer state among the impurities and a bath atom in Section~\ref{sec:3bd}. We summarize our findings and discuss future perspectives in Section~\ref{sec:conclusion}. The behavior of the logarithmic negativity in order to quantify the bipartite intercomponent entanglement is discussed in Appendix~\ref{ap:entanglement}. Appendices~\ref{ap:1dfit} and \ref{ap:2bfit_Casimir} provide supplemental information regarding the polaron characteristics and induced effective interactions. In Appendix~\ref{ap:mB2_NA30} we comment on the impact of the impurity mass and the number of bath particles on the ground state properties of the system. Finally, in Appendix~\ref{ap:nstate} we elaborate on the microscopic excitation processes of the system via a number state analysis.

\section{Two distinguishable impurities in a bosonic gas}
\label{sec:setup}

We consider a one-dimensional harmonically trapped three component mixture. It contains a bosonic medium $A$ with $N_A=15$ atoms of mass $m_A$ and two distinguishable impurities $B$ and $C$, i.e., $N_B=N_C=1$, having masses $m_B$ and $m_C$, respectively. The many-body Hamiltonian of this system reads
\begin{align}
    \hat{H} = \sum_{\sigma} \hat{H}_\sigma + \sum_{\sigma\neq\sigma'} \hat{H}_{\sigma\sigma'},
    \label{eq:Hamiltionian}
\end{align}
where $\hat{H}_\sigma$ denotes the Hamiltonian of each component $\sigma$ and $\hat{H}_{\sigma \sigma'}$ represents the intercomponent interaction contribution with $\sigma,\sigma'\in\{A,B,C\}$. 
Specifically, 
\begin{align}
    \hat{H}_\sigma &= \sum_{i=1}^{N_\sigma}
    \bigg( - \frac{\hbar^2}{2m_\sigma} \frac{\partial^2}{(\partial x_i^{\sigma})^2} +  \frac{1}{2}m_\sigma\omega_\sigma^2(x_i^\sigma)^2 + g_{\sigma\sigma} \sum_{i<j} \delta(x_i^\sigma-x_j^\sigma)\bigg),
    \label{eq:species_hamilt} \\
    \hat{H}_{\sigma\sigma'} &= g_{\sigma\sigma'} \sum_{i=1}^{N_\sigma} \sum_{j=1}^{N_{\sigma'}}\delta(x_i^\sigma-x_j^{\sigma'}).
\end{align}
Assuming that the system is at ultracold temperatures it dominantly experiences $s$-wave scattering processes that can be described by two-body contact interactions between particles of the same as well as of different species characterized by the generic strength $g_{\sigma\sigma'}$~\cite{chin2010}. 
The latter depends on the respective three-dimensional scattering lengths $a^{3D}_{\sigma\sigma'}$ and the transversal confinement frequency $\omega_{\perp}$ that are experimentally tunable via Feshbach resonances \cite{kohler2006, chin2010} and confinement induced resonances respectively \cite{olshanii1998}. The latter would allow the tuning of interactions even in the absence of a Feshbach resonance.

For simplicity, we focus on the mass-balanced case $m_\sigma \equiv m$ (unless stated otherwise) and thus $\omega_\sigma \equiv \omega$. Moreover, we rescale our Hamiltonian in harmonic oscillator units $\hbar\omega$ which means that the length and interaction scales are given in $\sqrt{\hbar/m\omega}$ and $\sqrt{\hbar^3\omega/m}$, respectively. Such a three-component system could be experimentally realized~\cite{taglieber2008, wu2011} e.g., by trapping three different hyperfine states of $^{87}$Rb which can feature various Feshbach resonances. 
An alternative candidate may be two isotopes of Rubidium atoms with $~^{85}$Rb emulating the medium and two-hyperfine states of $~^{87}$Rb~\cite{egorov2013, alvarez2013} representing the impurities.
Since our main findings persist also for mass-imbalanced mixtures, see the discussion in Section~\ref{ap:mB2_NA30}, corresponding heteronuclear mixtures of different isotopes could also be used. 
We also note that the experimental realization of three-component mixtures was reported in Refs.~\cite{taglieber2008,wu2011} and a proposal for a corresponding impurity system was recently made in Ref.~\cite{bighin2022impurity}.
Since our aim is to understand the role of induced interactions between the impurities mediated by the medium, in the ground state of the system, it is natural to consider two non-interacting impurities setting $g_{BC}=0$, which could be realized, for instance, via magnetic Feshbach resonances \cite{roati2008}.

\section{Variational wave function approach} \label{sec:mlx-method}

The ground state of the three-component mixture, described by the Hamiltonian of Eq.~(\ref{eq:Hamiltionian}), is determined within the ML-MCTDHX method~\cite{kronke2013, cao2013, cao2017, kohler2019}. A central aspect of this \textit{ab-initio} approach is based on the expansion of the many-body wave function on different layers using a variationally optimized time-dependent many-body basis. This leads to an efficient truncation of the underlying Hilbert space tailored to capture the relevant inter- and intracomponent correlations. Specifically, the many-body wave function is first expressed in terms of three different sets of $D_{\sigma}$ species functions as follows 
\begin{align}
\ket{\Psi^{\mathrm{MB}}(t)} = \sum_{i=1}^{D_A}\sum_{j=1}^{D_B}\sum_{k=1}^{D_C} C_{ijk}(t) \ket{\Psi_i^A(t)} \ket{\Psi_j^B(t)} \ket{\Psi_k^C(t)}.
\label{eq:wavefct_toplayer}
\end{align} 
The time-dependent coefficients $C_{ijk}(t)$ bare information about the entanglement between the involved components. For instance, the bipartite entanglement between two components can be analyzed by tracing out the degrees of freedom of the third one and then apply the positive partial transpose criterion on the resulting mixed state~\cite{zyczkowski1998} (see also Appendix~\ref{ap:entanglement}). 
Next, the intracomponent correlations are included into the wave function ansatz by expanding each species function as a superposition of permanents $\ket{\vec{n}(t)}$ weighted by time-dependent expansion coefficients $C_{i,\vec{n}}^{\sigma}(t)$. In this notation, $\vec{n}=(n_1^\sigma,\dots, n_{d^\sigma})$ represents the occupation distribution of $N_\sigma$ particles on $d_\sigma$ time-dependent single-particle functions. Additionally, the single-particle functions are expanded into a time-independent discrete variable representation~\cite{light1985} consisting in our case of $\mathcal{M}_r=300$ evenly spaced grid points.

The number of utilized species functions $D_\sigma$ dictates the degree of intercomponent entanglement. For instance, by providing only one species function for each component, i.e., by setting $D_A=D_B=D_C=1$, the many-body wave function reduces on its top layer to a product state, thereby, prohibiting any interspecies entanglement. Such a treatment is commonly referred to as a species mean-field ansatz (sMF)~\cite{kronke2013}. For two-component mixtures the sMF ansatz is unique, however, in three-component systems there are various sMF that could be constructed. As an example, setting $D_\sigma=1$ and $D_{\sigma'}, D_{\sigma''} > 1$, we allow for entanglement generation only between the species $\sigma'$ and $\sigma''$, whilst intercomponent correlations with species $\sigma$ are suppressed. To clearly distinguish among the different possible sMF ansatzes, in the following, we abbreviate as sMF$\sigma$ where $\sigma\in \{A,B,C\}$ the ansatz that ignores intercomponent correlations between species $\sigma$ and the remaining ones. In this sense, the sMFC is written as 
\begin{align}
    \ket{\Psi^{\mathrm{sMFC}}(t)} = \sum_{i=1}^{D_A}\sum_{j=1}^{D_B} C_{ij1}(t) \ket{\Psi_i^A(t)} \ket{\Psi_j^B(t)} \ket{\Psi_1^C(t)},
    \label{eq:sMFC_ansatz}
\end{align}
where only species $A$ and $B$ can become entangled while species $C$ remains uncorrelated with the other species.

The ground state of the three component mixture is obtained through the imaginary time propagation method. The time-dependent coefficients of each layer, namely the species and single-particle layers, are optimally adapted to the system, e.g. by following the Dirac-Frenkel variational principle~\cite{raab2000} in order to determine the underlying ML-MCTDHX equations of motion. 
The latter correspond to $D_AD_BD_C$ linear differential equations of motion for the $C_{ijk}(t)$ coefficients coupled to $\sum_{\sigma=A,B,C} D_\sigma {{N_\sigma + d_\sigma -1}\choose{d_\sigma - 1}}$ nonlinear integrodifferential equations for the species functions and $d_A+d_B+d_C$ nonlinear integrodifferential equations for the single-particle functions. 
This co-moving basis concept minimizes the number of required states for achieving numerical convergence. In this sense, it reduces the computational cost as compared to methods relying on time-independent basis sets, while simultaneously allows to account for all relevant correlations. The truncation of the Hilbert space is determined by the number of employed species- and single-particle functions defining the numerical configuration space ($D_A$, $D_B$, $D_C$; $d_A$, $d_B$, $d_C$).
Utilizing this method, it is in principle possible to describe mixtures with mesoscopic particle numbers and strong interactions. However, as the number of particles increases and correlations become enhanced a larger number of orbitals should be taken into account in order to reach numerical  convergence. The latter is carefully checked by systematically increasing the numerical configuration space and ensuring that the observables of interest remain unchanged within a desired level of accuracy. As expected, this process is accompanied by a significant computational cost and in particular it is the interplay of intra- and intercomponent correlations with the components atom number that limits the applicability of the method due to numerical feasibility. 
Elaborated discussions on the ingredients, applicability and benchmarks of this variational method to different multicomponent settings can be found in the recent reviews~\cite{mistakidis2022cold,lode2020colloquium}.

For our system, the degree of correlations in the bosonic bath, e.g. as captured by its depletion~\cite{penrose1956} $1-n^A_0$ with $n_0^A$ representing the largest eigenvalue of the bath's one-body reduced density matrix is negligible within the considered interaction strength intervals. This allows us to use only a few orbitals for the medium in order to ensure convergence. On the other hand, the impurities depletion is in general larger, especially for strongly repulsive interactions, and thus we need to use more orbitals. Herewith, we have checked that employing an orbital configuration (6, 6, 6; 4, 6, 6) results in the convergence of the observables of interest, such as the species densities and intercomponent two-body correlation functions, while the amount of equations of motion are tractable. For completeness, let us note that stronger intercomponent interactions than the ones to be reported below e.g. $\abs{g_{AC}}<10$ require a larger number of species functions and impurities orbitals which is still numerically feasible. Similarly, in order to tackle systems with stronger intracomponent bath interactions the number of the respective $d_A$ orbitals should be increased. This naturally entails more difficult convergence issues than increasing the impurities orbitals (and thus considering stronger impurity-medium interactions) since the number of the underlying equations of motion becomes larger in the former case.

\section{One-body density configurations of the three-component mixture}
\label{sec:1bd}

\begin{figure}[t]
    \centering
    \includegraphics[width=0.9\linewidth]{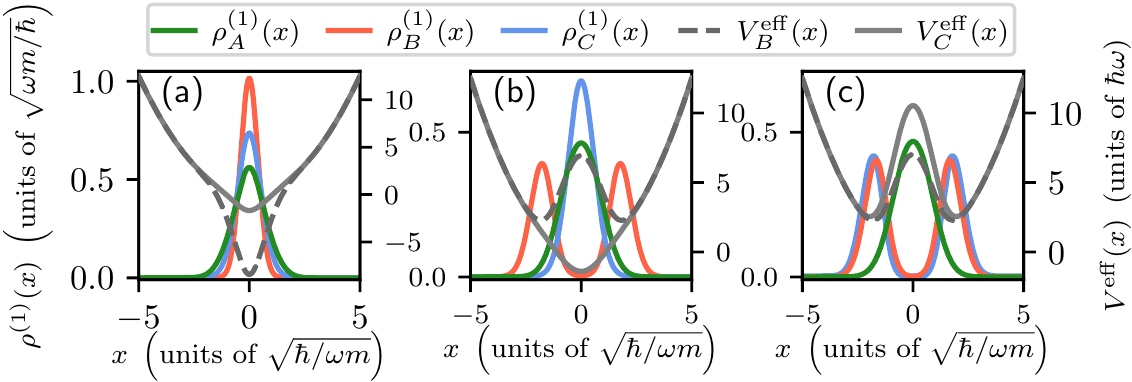}
    \caption{One-body $\sigma$-species density, $\rho_\sigma^{(1)}(x)$, shown together with the effective potentials [Eq. (\ref{eq:effpot})] of the impurities (see legend). Two distinguishable non-interacting impurities ($B$, $C$) are considered which are individually coupled to a bosonic medium $A$ with $g_{AA}=0.2$. The impurity-medium coupling strengths from left to right panels refer to $(g_{AB}, g_{AC})=(-1.0, -0.2)$, $(1.0, -0.2)$ and $(1.0, 1.5)$. For attractive interactions the medium atoms accumulate in the vicinity of the impurities and their effective potential is attractive. Turning to repulsive couplings a tendency for impurity-medium phase-separation occurs for $g_{A\sigma}>g_{AA}$.}
\label{fig:gpop}
\end{figure}

To investigate the emergent spatial configurations of the three-component impurity setting arising due to different combinations of the involved interactions, we initially employ the $\sigma$-component one-body density being normalized to unity. 
Namely, $\rho_\sigma^{(1)}(x) = \bra{\Psi^{\mathrm{MB}}} \hat{\Psi}_\sigma^\dagger(x) \hat{\Psi}_\sigma(x)$ $ \ket{\Psi^{\mathrm{MB}}}$ where $\hat{\Psi}_\sigma^{(\dagger)}$ denotes the bosonic field operator which annihilates (creates) a $\sigma$-species atom at position $x$. In an experiment, the density is routinely detected through \textit{in-situ} absorption imaging \cite{catani2009,cheiney2018,mistakidis2018correlation}. Our understanding on the mixture spatial distributions at different interactions is also corroborated by an effective potential picture, which has been proven thus far successful in order to qualitative explicate various aspects of impurity physics in two-component settings~\cite{mistakidis2019a,mistakidis2020b,theel2021}. According to this, each $\sigma$ component is subjected to an effective potential stemming from the superposition of its external harmonic trap and the density of the complementary components $\sigma'$ weighted by the respective intercomponent interactions, i.e.,
\begin{align}
    V_\sigma^{\mathrm{eff}}(x) = V_\sigma(x) + \sum_{\sigma' \neq \sigma} N_{\sigma'} g_{\sigma\sigma'} \rho_{\sigma'}^{(1)}(x).
    \label{eq:effpot}
\end{align}
Naturally, this is a sMF framework since it ignores intercomponent correlations. Moreover, it is more meaningful for the impurity subsystem since the impact of the impurity densities is suppressed for the medium. Density profiles of all three components and the impurity effective potentials are provided in Fig.~\ref{fig:gpop} for characteristic impurity-medium interaction configurations, namely $(g_{AB}, g_{AC})=(-1.0, -0.2)$, $(1.0, -0.2)$ and $(1.0, 1.5)$. 
The impurities are considered to be non-interacting among each other, i.e., $g_{BC}=0$, and the medium bosons feature throughout $g_{AA}=0.2$. 

As it can be seen, for an overall attractive impurity-medium coupling the bosons of the medium are placed in the vicinity of the impurities which are naturally localized at the trap center [cf. Figure \ref{fig:gpop}(a)]. This distribution of the medium atoms can also be understood in terms of the respective attractive impurity-medium interaction energy $E_{A\sigma}^{\mathrm{int}}=\langle \Psi^{\mathrm{MB}}| \mathcal{H}_{A \sigma} |\Psi^{\mathrm{MB}} \rangle$ for $g_{A\sigma}<0$ with $\sigma=B,C$. Also, for both $g_{AB}<0$ and $g_{AC}<0$ the effective potential of each impurity corresponds to a dipped harmonic trap enforcing its localization whose degree is, of course, enhanced for stronger attractions [cf. Figure \ref{fig:gpop}(a)]. The value of the attractive interaction determines the degree of spatial localization, i.e., the $B$ impurity with $g_{AB}=-1.0$ is more localized than the $C$ impurity experiencing $g_{AC}=-0.2$. For sufficiently large attractive impurity-medium couplings ($\abs{g_{A\sigma}} \gg g_{AA}$) the impurities form a bipolaron, see for details the discussion in Section~\ref{sec:bip}.

On the other hand, tuning at least one of the impurity-medium couplings towards the repulsive regime such that $g_{A \sigma}>g_{AA}$ is satisfied leads to the phase-separation among these components since $E_{A\sigma}^{\mathrm{int}}>0$. In this case, the impurity forms a shell around the edges of the bath residing around the trap center~\cite{keiler2021}. Such configurations can be readily observed, for instance, in Figure~\ref{fig:gpop}(b) where solely the $B$ impurity is strongly repulsively coupled with the bath ($g_{AB}>g_{AA}$) and also in Figure~\ref{fig:gpop}(c) where both impurities phase separate with the bath due to $g_{AB}>g_{AA}$ and $g_{AC}>g_{AA}$. Notice that for strong repulsive impurity-medium couplings the underlying effective potential of the impurity has the form of a double-well potential which favors the phase-separation among the bath and the corresponding impurity [cf. Figures \ref{fig:gpop}(b) and (c)]. 

Another interesting phenomenon reflecting the richness of three-component systems arises upon considering distinct interactions between each impurity and the bath. Indeed, varying the impurity-medium coupling for a specific impurity affects the shape of the bath accordingly and, in turn, impacts the distribution of the other impurity. This is visualized in Figures~\ref{fig:gpop}(a) and (b) where $g_{AC}$ is the same while $g_{AB}$ is modified from attractive to repulsive values ultimately altering the spatial localization of impurity $C$, see in particular the peak of $\rho_C^{(1)}(x)$. Therefore, it is possible to implicitly manipulate the distribution of one impurity by adjusting the coupling of the other impurity with the bath and importantly in the absence of direct impurity-impurity interaction. This property, as it will be discussed below, can be proved crucial for controlling impurity-impurity induced interactions.

\begin{figure}[t]
\centering
\includegraphics[width=0.85\linewidth]{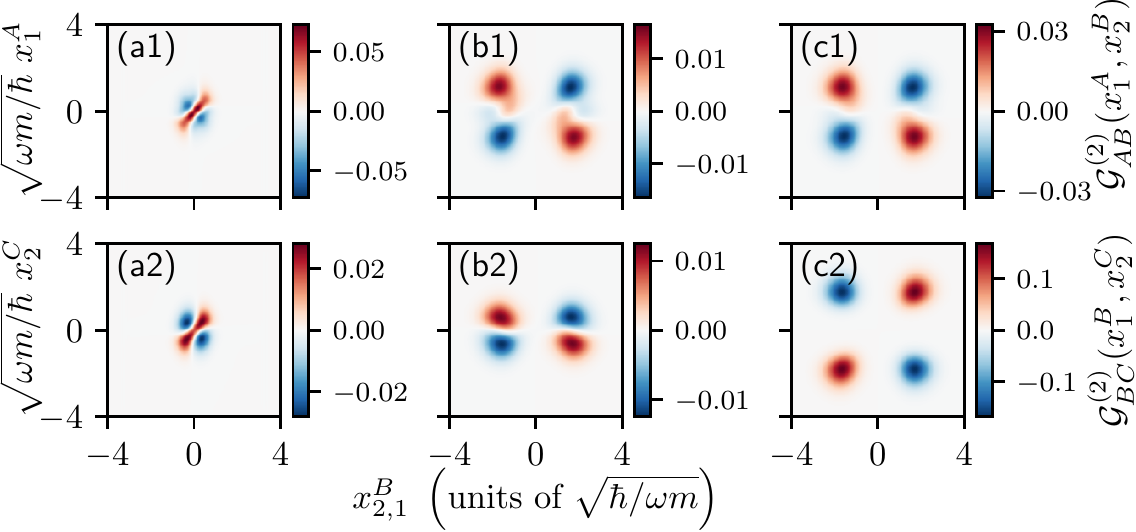}
\caption{Two-body correlation function (in units of $ m\omega/\hbar$) between (a1)-(c1) one bath particle and the $B$ impurity as well as (a2)-(c2) among the two non-interacting impurities [see Eq.~(\ref{eq:noise_correlation})].
Each column corresponds to the same interaction configuration which is from left to right $(g_{AB}, g_{AC})=(-1.0, -0.2)$, $(1.0, -0.2)$ and $(1.0, 1.5)$. We consider two distinguishable non-interacting impurities and an interacting medium with $g_{AA}=0.2$. Impurity $B$ is correlated (anti-correlated) with a bath particle at the same location in the case of attractive (repulsive) $g_{AB}$, see panel (a1) [(b1), (c1)]. The impurities experience induced correlations when they both couple either repulsively or attractively to the bath [panels (a2), (c2)], while they are anti-correlated when each impurity couples with an opposite sign to the majority species [panel (b2)].} 
    \label{fig:corr}
\end{figure}

\section{Intercomponent (induced) correlations and entanglement}
\label{sec:corr}

Next, we shed light on the associated intercomponent correlation patterns with a particular emphasis on the existence of induced correlations between the impurities mediated by the bosonic gas. The intercomponent two-body spatial correlations, or two-body coherence, can be quantified through~\cite{mistakidis2018correlation},
\begin{equation}
    \mathcal{G}_{\sigma\sigma'}^{(2)}(x_1^\sigma, x_2^{\sigma'}) = \rho_{\sigma\sigma'}^{(2)}(x_1^\sigma, x_2^{\sigma'}) - \rho_{\sigma}^{(1)}(x_1^\sigma)\rho_{\sigma'}^{(1)}(x_2^{\sigma'}).
    \label{eq:noise_correlation}
\end{equation}
Here, we subtract the probability of independently detecting a $\sigma$ and a $\sigma'$ atom at positions $x_1^\sigma$ and $x_2^{\sigma'}$ from the probability to simultaneously measure one at $x_1^\sigma$ and the other at $x_2^{\sigma'}$. The latter is provided by the reduced two-body density
\begin{equation}
    \rho_{\sigma\sigma'}^{(2)}(x_1^\sigma, x_2^{\sigma'}) = \bra{\Psi^{\mathrm{MB}}} \hat{\Psi}_\sigma^\dagger(x_1^\sigma) \hat{\Psi}_{\sigma'}^\dagger(x_2^{\sigma'}) \hat{\Psi}_{\sigma'}(x_2^{\sigma'}) \hat{\Psi}_\sigma(x_1^\sigma) \ket{\Psi^{\mathrm{MB}}},
    \label{eq:2bd}
\end{equation} 
which is normalized to unity. 
In this sense, the two particles are correlated or bunched (anti-correlated or antibunched) if $\mathcal{G}_{\sigma\sigma'}^{(2)}(x_1^\sigma, x_2^{\sigma'})$ is positive (negative); otherwise, they are referred to as two-body un-correlated~\cite{mistakidis2018correlation,keiler2020}. Experimentally the two-body correlation function is accessible through analyzing the respective single-shot images, see e.g. Refs.~\cite{hodgman2011, dall2013, nguyen2019, borselli2021, hofferberth2008}.

\subsection{Characteristic correlation patterns}
\label{sec:corr_patterns}

First, we study the emergent two-body correlation patterns between the $B$ impurity and the medium for different intercomponent interactions [Figures~\ref{fig:corr}(a1)-(c1)]. For attractive $g_{AB}<0$ and $g_{AC}<0$ the $B$ impurity is correlated with a bath atom at the same position, see the diagonal of $\mathcal{G}_{AB}^{(2)}(x_1^A, x_2^B)>0$, while these two particles are anti-correlated when symmetrically placed with respect to the trap center as it is shown from the anti-diagonal of $\mathcal{G}_{AB}^{(2)}(x_1^A, x_2^B)<0$ [Figure~\ref{fig:corr}(a1)]. In this sense, the $B$ impurity prefers to occupy the same spatial region with the bath. Turning to repulsive $g_{AB}>0$ and independently of $g_{AC}\lessgtr 0$, the above-discussed two-body correlation distributions are inverted and the $B$ impurity features an anti-bunched (bunched) behavior at the same (different) location with a bath particle as can be deduced by the diagonal (anti-diagonal) of $\mathcal{G}_{AB}^{(2)}(x_1^A, x_2^B)$ [cf. Figures~\ref{fig:corr}(b1) and (c1)]. 
This trend reflects the impurity-medium phase-separation identified on the density level [Figures~\ref{fig:gpop}(b) and (c)].

Let us now discuss the induced correlations among the non-interacting impurities. When both impurities are attractively coupled to their bath they exhibit a bunching tendency which is, of course, mediated by the bosonic gas, see the diagonal of $\mathcal{G}_{BC}^{(2)}(x_1^B, x_2^C)$ depicted in Figure~\ref{fig:corr}(a2). Otherwise, the impurities are anti-bunched when residing at different locations with respect to $x=0$. This two-body configuration of the impurities manifests the presence of their attractive induced interactions regulated by the impurity-medium attractive interactions as we will discuss in Section~\ref{sec:induced_interactions}.
Note also that a further increase of the impurity-bath attraction can result in the formation of a bipolaron state which we analyze in detail within Section \ref{sec:bip}. A similar two-body impurities correlation pattern occurs when they both repulsively couple with their bath [Figure~\ref{fig:corr}(c2)]. However, in this case the impurities cluster either at the left or the right side of the bath, while the probability to reside at opposite sides is suppressed [cf. Figure \ref{fig:corr}(c2)]. This trend which is inherently related to the impurity-medium phase-separation has also been observed for two indistinguishable impurities and it is known as their coalescence~\cite{dehkharghani2018}. In sharp contrast, if one impurity couples repulsively and the other attractively to the bath the reverse to the above-described correlation behavior is observed. 
Namely, the impurities anti-bunch (bunch) at the same (different) location in terms of the trap center, see Figure \ref{fig:corr}(b2). This scenario manifests the flexibility offered by three component mixtures and it is connected to the emergence of repulsive impurity-impurity induced interactions, a phenomenon that can not occur in two-component systems and we analyze in Section~\ref{sec:induced_interactions}.

\begin{figure*}[t]
    \centering
    \includegraphics[width=0.75\linewidth]{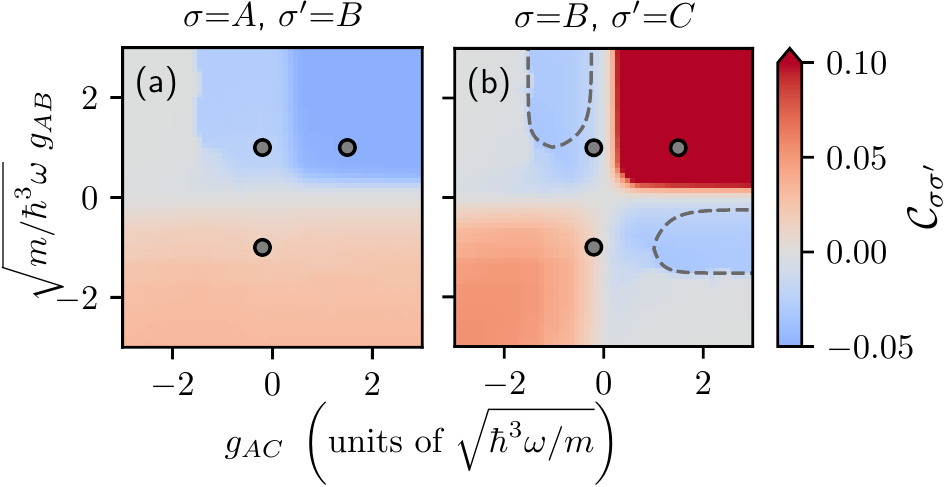}
    \caption{(a)-(b) Phase diagram of the intercomponent (see legends) spatially integrated correlation functions $\mathcal{C}_{\sigma\sigma'}$ [Eq.~(\ref{eq:int_noise_corr})] in the parametric plane of the impurity-medium interaction strengths ($g_{AB}$, $g_{AC}$). A value of $\mathcal{C}_{\sigma\sigma'} < 0$ ($\mathcal{C}_{\sigma\sigma'} > 0$) indicates an anti-correlated (correlated) behavior between the atoms of species $\sigma$ and $\sigma'$, while $\mathcal{C}_{\sigma\sigma'}=0$ denotes the absence of two-body correlations (see also main text). The gray circles correspond to the interaction combinations ($g_{AB}$, $g_{AC}$) depicted in Figures~\ref{fig:gpop} and \ref{fig:corr}. The regions enclosed by the dashed lines in panel (b) indicate the interaction regions where the impurities do not overlap but are still two-body anti-correlated. The harmonically trapped three component system consists of two non-interacting but distinguishable impurities immersed in a bosonic gas of $N_A=15$ atoms with $g_{AA}=0.2$.}
    \label{fig:phase_diag}
\end{figure*}

\subsection{Emergent correlation regimes}
\label{sec:corr_regime}

To provide an overview of the two-body correlation behavior stemming from the interplay of the distinct impurity-medium couplings, we inspect the spatially integrated over $\ [ -\infty,0 \ ]$ (due to symmetry) correlation function
\begin{align}
    \mathcal{C}_{\sigma\sigma'} = \int_{-\infty}^{0} dx_1^\sigma \int_{-\infty}^{0} dx_2^{\sigma'} \mathcal{G}_{\sigma\sigma'}^{(2)}(x_1^\sigma, x_2^{\sigma'}).
    \label{eq:int_noise_corr}
\end{align}
It quantifies the amount of intercomponent correlations or anti-correlations by means that it is positive (negative) when the particles prefer (avoid) to occupy the same region with respect to the trap center\footnote{Due to parity symmetry the maximum (minimum) value of $\mathcal{C}_{\sigma\sigma'}$ is 0.25 (-0.25) denoting strong bunching (anti-bunching).}. The phase diagrams of the impurity-medium $\mathcal{C}_{AB}$ and impurity-impurity $\mathcal{C}_{BC}$ integrated correlations as a function of $g_{AB}$ and $g_{AC}$ are depicted in Figure \ref{fig:phase_diag}(a) and (b) respectively. Recall that since $g_{BC}=0$ all emerging impurity correlations are induced by their coupling to the bath. 

An anti-correlated (correlated) behavior between the $B$ impurity and the bath occurs for $g_{AB}>0$ ($g_{AB}<0$) and varying $g_{AC}$, see also Figures \ref{fig:corr}(a1)-(c1). Notice also the un-correlated tendency for strongly attractive $g_{AC}$ and repulsive $g_{AB}$ [Figures \ref{fig:phase_diag}(a), (b)]. 
Indeed, due to the large $g_{AC}<0$ both the bath $A$ and the $C$ impurity localize at the trap center minimizing their spatial overlap with the $B$ impurity since $g_{AB}>0$ and thus $\mathcal{C}_{AB}$ is suppressed. Naturally, a less attractive $g_{AC}$ enhances the overlap between impurity $B$ and the bath leading to an anti-correlated behavior. The largest degree of anti-correlation as captured by $\mathcal{C}_{AB}$ is reached when $g_{AB}>g_{AA}$ and $g_{AC}>g_{AA}$ where both impurities form a shell around the bath and coalesce [cf. corresponding region in Figure \ref{fig:phase_diag}(a)].

Turning to the impurities' correlations, we observe that as long as they both couple either repulsively or attractively to the bath it holds that $\mathcal{C}_{BC}>0$, implying that they are correlated [see also Figures~\ref{fig:corr}(a1) and (c1)]. However, when the couplings $g_{AB}$ and $g_{AC}$ have opposite signs, with one lying in the weak and the other in the strong interaction regime, then mostly $\mathcal{C}_{BC}<0$, i.e., the impurities are anti-correlated [cf. Figure \ref{fig:corr}(b1)]. A notable exception takes place if one of the impurities couples strongly repulsively to the bath (e.g. $g_{AB}>g_{AA}$) and the other strongly attractively (e.g. $\abs{g_{AC}}>g_{AA}$). This leads to a suppressed spatial overlap among the bath and the repulsively interacting impurity\footnote{Notice here that since the impurity $B$ is neither entangled with the bath nor with the impurity $C$, it is sufficient to consider the sMFB ansatz. We have checked that $|\langle \Psi^{\mathrm{sMFB}} | \Psi^{\mathrm{MB}} \rangle|^2\approx 1$ holds, see also Appendix~\ref{ap:nstate} for a detailed number state analysis of the many-body wave function.} and thus the bath is only correlated with the attractively coupled impurity, see also the discussion above. Together with the fact that the impurities are spatially separated in this interaction region, if mediated impurity correlations occur they have to be nonlocal. This is indeed the case since the impurities are found to be anti-correlated, $\mathcal{C}_{BC}<0$, see the two parameter regimes in Figure~\ref{fig:phase_diag}(b) enclosed by the dashed lines.

\section{Quantification of impurities induced interactions}
\label{sec:induced_interactions}

Below, we examine how the mediated correlations among the distinguishable impurities alter their relative distance and, subsequently, relate the induced impurity-impurity correlation patterns with an effective induced interaction strength. The latter as it will be argued can be either attractive or repulsive due to the genuine three-component nature of the system and it is further quantified via an effective two-body model. 

\subsection{Effect of the induced impurity-impurity correlations on their relative distance}
\label{sec:induced_interactions_rel_dist}

A reliable measure for this purpose, that has also been utilized in two-component settings \cite{keiler2020,mistakidis2020b} and can be experimentally monitored via \textit{in-situ} spin-resolved single-shot measurements \cite{bergschneider2018}, is the relative distance between the impurities 
\begin{align}
    \langle r_{BC} \rangle = \frac{1}{N_BN_C}\int {\mathrm{d}} x_1^B {\mathrm{d}} x_2^C \left|x_1^B - x_2^C \right| \rho_{BC}^{(2)}(x_1^B, x_2^C).
    \label{eq:rel_dist}
\end{align}
Specifically, in order to extract the contribution stemming from genuine impurity-medium correlations we estimate the modified relative distance at different correlation levels as dictated by the respective truncation of the many-body (MB) wave function (see also Section~\ref{sec:mlx-method}), namely
\begin{equation}
    \Delta \langle r_{BC} \rangle =
    \langle r_{BC}^{\mathrm{MB}} \rangle - \left[
    \langle r_{BC}^{\mathrm{sMF}} \rangle
    + \left(\langle r_{BC}^{\mathrm{sMFB}} \rangle - \langle r_{BC}^{\mathrm{sMF}} \rangle \right)
    + \left(\langle r_{BC}^{\mathrm{sMFC}} \rangle - \langle r_{BC}^{\mathrm{sMF}} \rangle \right)  \right].
    \label{eq:rel_dist_diff}
\end{equation}
Here, sMF stands for the general species mean-field case where all intercomponent correlations are neglected, while sMFB (sMFC) refers to the case at which only intercomponent correlations between the $B$ ($C$) impurity and the medium are ignored~\cite{keiler2021,mistakidis2022cold}. 
Excluding the sMF contribution as well as the ones corresponding to the entanglement between the bath and impurity $C$ or $B$ [cf. last four terms of Eq.~(\ref{eq:rel_dist_diff})] from the relative distance where all correlations are included, i.e., $\langle r_{BC}^{\mathrm{MB}} \rangle$, we are able to distill the effects originating from the mutual correlation among the impurities and the bosonic gas by tracking $\Delta \langle r_{BC} \rangle$. 
As such, $\Delta \langle r_{BC} \rangle$ captures the genuine effects of the induced correlations as described by $\mathcal{C}_{BC}$ [Figure~\ref{fig:phase_diag}(b)]. 
We interpret a value of $\Delta \langle r_{BC} \rangle$ which is positive (negative) as the signal of emergent repulsive (attractive) impurities' induced interactions.

\begin{figure*}
    \centering
    \includegraphics[width=1.0\linewidth]{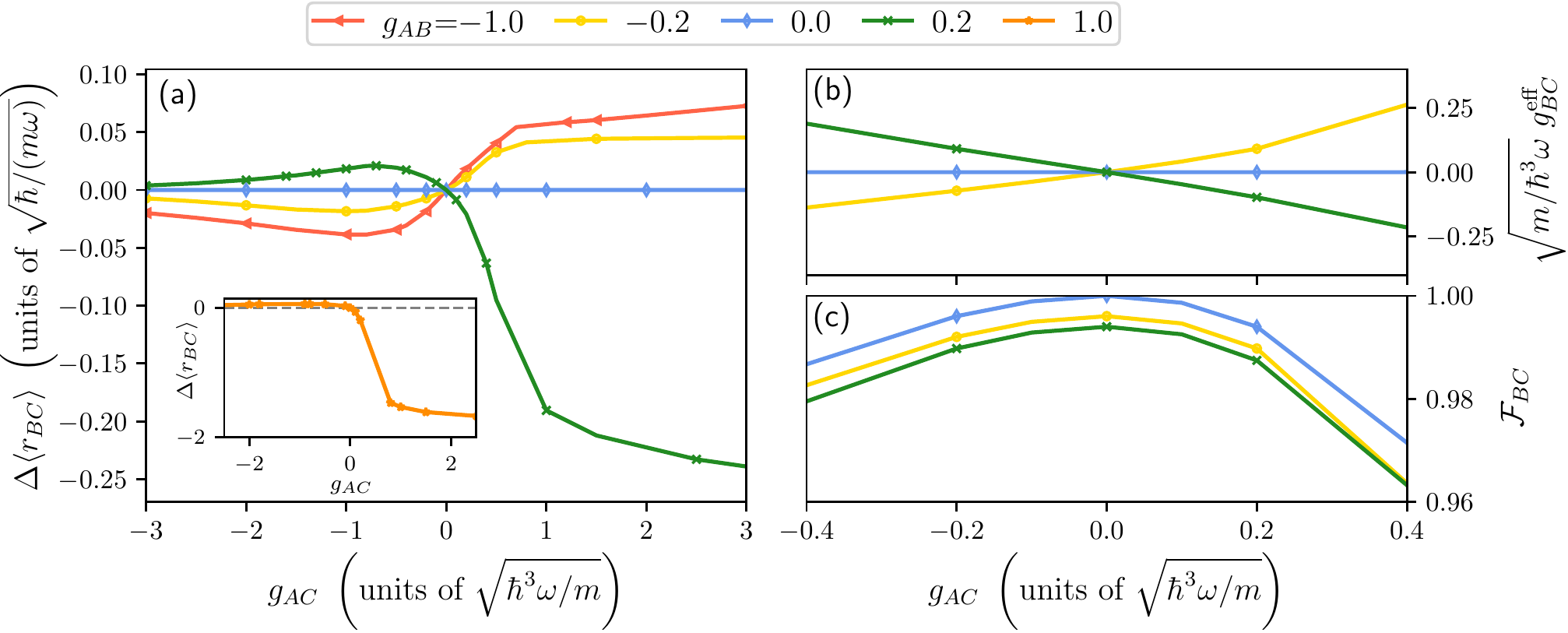}
    \caption{(a) and its inset: Modified relative distance [Eq.~(\ref{eq:rel_dist_diff})] reflecting the effects on $\langle r_{BC}^{\mathrm{MB}} \rangle$ which are exclusively caused by the induced impurities correlation as a function of $g_{AC}$ and for different fixed $g_{AB}$.
    (b) Induced interaction strength between the two Bose polarons estimated by maximizing the overlap between the two-body correlation functions $\mathcal{G}_{BC}^{(2), \mathrm{eff}}$ obtained from the effective two-body model and $\mathcal{G}_{BC}^{(2)}$ predicted within the many-body approach (see main text). (c) Fidelity $\mathcal{F}_{BC}$ of the impurities wave function as found in the many-body method and the effective two-body model with respect to the impurity-medium couplings $g_{AB}$ and $g_{AC}$.
    We consider two non-interacting but distinguishable impurities immersed in a bosonic gas of $N_A=15$ atoms with $g_{AA}=0.2$.} 
    \label{fig:induced_interaction}
\end{figure*}

The modified relative distance, $\Delta \langle r_{BC} \rangle$, is presented in Figure~\ref{fig:induced_interaction}(a) with respect to the $g_{AC}$ coupling and for characteristic fixed $g_{AB}$ values. In general, we find an induced attraction between the impurities when they both couple either attractively or repulsively to the medium, while they feature a mediated repulsion if one of them couples attractively and the other repulsively to the bosonic gas. 
Since $\Delta \langle r_{BC} \rangle$ is closely related to $\mathcal{C}_{BC}$, an induced correlation (anti-correlation) between the impurities can be associated to their attractive (repulsive) induced interaction and vice versa [cf. Figures~\ref{fig:phase_diag}(b) and \ref{fig:induced_interaction}(a)].
For instance, considering repulsive $g_{AB}$ and tuning $g_{AC}$ to weak attractions, $\Delta \langle r_{BC} \rangle$ becomes positive denoting an induced repulsion between the impurities. However, for stronger repulsive $g_{AC}$ $\Delta \langle r_{BC} \rangle$ is negative and thus attractive induced interactions occur maximizing in the coalescence regime where $g_{AB}$ and $g_{AC}$ are both strongly repulsive, see also the inset of Figure~\ref{fig:phase_diag}(a). 
Furthermore, in the case of suppressed mediated correlations between the impurities ($\mathcal{C}_{BC}\approx0$), i.e., in the trivial case where $g_{AB}=0$ or for strong attractive $g_{AC}$ and repulsive $g_{AB}$ [cf. Figure~\ref{fig:phase_diag}(b)], also $\Delta \langle r_{BC} \rangle$ vanishes (see Figure \ref{fig:induced_interaction}(a) for strong attractive $g_{AC}$ and $g_{AB}=0.2,1.0$). In the last scenario, the gradually increasing $g_{AC}$ attraction leads to a reduction (enhancement) of the correlation between the bath and the $B$ ($C$) impurity whose interplay impedes the development of mediated impurity correlations and therefore induced interactions.

In the case of an attractively coupled impurity $B$, e.g. $g_{AB}=-1.0, -0.2$, $\Delta \langle r_{BC} \rangle$ decreases when $g_{AB}$ is tuned to strong attractive values, a phenomenon also occurring for $\mathcal{C}_{BC}$ [Figure~\ref{fig:phase_diag}(b)]. Here, increasing the attraction between impurity $C$ and the bath enhances their correlation, while at sufficiently strong attractive $g_{AC}$ the correlation between the bath and the impurity $B$ begins to slightly decrease for constant attractive $g_{AB}$ (cf. Figure~\ref{fig:phase_diag}). This competition between the different impurity-medium correlations suggests an interesting interplay between the individual intercomponent correlations and could in principle hinder the bath to mediate correlations between the impurities leading eventually to the observed reduction of the induced impurity-impurity correlation/interaction. Such an interplay of intercomponent correlations is indicative of a more intricate and generic correlation transfer process among the species~\cite{mistakidis2022cold}, that is an exciting future perspective but lies beyond the focus of our study. However, note that for decreasing $g_{AB}=g_{AC}$ results in a saturation of the impurity-impurity correlation, a fact that will also become important later in the discussion regarding the bipolaron formation in Section~\ref{sec:bip}.

Finally, notice that a similar qualitative behavior of the intercomponent correlations and thus also of $\Delta \langle r_{BC} \rangle$ takes place for either increasing the number of atoms of the bosonic medium or the bare mass of one of the impurities, see Appendix~\ref{ap:mB2_NA30}. In fact, both scenarios lead for repulsive $g_{AB}$ and $g_{AC}$ to an amplified impurities entanglement and to a stronger attractive induced interaction.

\subsection{Effective two-body model}
\label{sec:induced_interactions_eff_2b_model}

To determine the strength of induced impurity-impurity interactions, we reduce the three-component many-body system to an effective two-body model consisting of two interacting quasi-particles. This is a common approach to identify polaron properties from many-body simulations and has been successfully applied to two indistinguishable impurities \cite{mistakidis2020a} but not to distinguishable ones. 
Here, the effective two-body model employs the effective potential $V_\sigma^{\mathrm{eff}}(x^\sigma)$ [defined in Eq.~(\ref{eq:effpot})] for each impurity and thus neglects impurity-medium correlations. Also, the underlying impurities induced interactions are represented by a contact potential of strength $g_{BC}^{\mathrm{eff}}$ (a treatment with finite range interactions leads to similar results as it is demonstrated in Appendix~\ref{ap:2bfit_Casimir}). 
Specifically, the corresponding effective two-body Hamiltonian reads
\begin{align}
    H^{(2),\mathrm{eff}} = \sum_{\sigma=B,C} \left(
     - \frac{\hbar^2}{2m_\sigma} \frac{\partial^2}{(\partial x^{\sigma})^2} + V_\sigma^{\mathrm{eff}}(x^\sigma) \right) + g_{BC}^{\mathrm{eff}} \delta(x^B-x^C).
    \label{eq:eff_two_body_model}
\end{align}
The effective potential accounts for the effective mass and frequency of each impurity~\cite{mistakidis2019}. These effective parameters originate from the polaron picture where the impurity becomes dressed by the excitations of the bath, see Appendix~\ref{ap:1dfit} for a more detailed discussion.

In order to deduce the effective interaction strength $g_{BC}^{\mathrm{eff}}$, we minimize $\Delta \mathcal{G}_{BC}^{(2)} = \int \mathrm{d}x_B\mathrm{d}x_C $ $\left| \mathcal{G}_{BC}^{(2)} - \mathcal{G}_{BC}^{(2), \mathrm{eff}} \right|^2$, where $\mathcal{G}_{BC}^{(2)}$ and $\mathcal{G}_{BC}^{(2), \mathrm{eff}}$ are the impurities' two-body correlation functions calculated from the many-body three-component mixture and the effective two-body model, respectively~\footnote{We find $\Delta \mathcal{G}_{BC}^{(2)}\lesssim 10^{-5}$ for all considered interaction strengths $g_{AB}$ and $g_{AC}$.}.
By estimating the value of $g_{BC}^{\mathrm{eff}}$ which minimizes $\Delta \mathcal{G}_{BC}^{(2)}$, we are able to associate the emergent induced correlation pattern between the impurities described in Fig.~\ref{fig:phase_diag}(b) with a corresponding induced interaction strength $g_{BC}^{\mathrm{eff}}$. 
The resultant behavior of $g_{BC}^{\mathrm{eff}}$ provided in Figure~\ref{fig:induced_interaction}(b) for fixed $g_{AB}$ and varying $g_{AC}$ agrees qualitatively with the observations made for $\Delta \langle r_{BC} \rangle$. The impurities experience an induced attraction when they both couple either attractively or repulsively to the bath, corresponding to an induced correlation, otherwise they feature an induced repulsion related to their anti-correlated tendency \footnote{
Note that $g_{BC}^{\mathrm{eff}}=0$ if one of the impurities does not interact with the bath which further confirms the validity of the effective model predictions since in this case no correlations are mediated.}.
To testify the validity range of the effective two-body model [Eq.~(\ref{eq:eff_two_body_model})] for describing the impurities, we calculate the fidelity $\mathcal{F}_{BC}$ of their ground state wave function as obtained from $H^{(2),\mathrm{eff}}$ ($\ket{\Phi_{\mathrm{eff}}^{BC}}$) and the full three-component mixture ($\ket{\tilde{\Psi}_i^{BC}}$)~\footnote{
For this reason we use the Schmidt decomposition $\ket{\Psi^{\mathrm{MB}}} = \sum_{i} \sqrt{\lambda_i} \ket{\tilde{\Psi}_i^A}\otimes \ket{\tilde{\Psi}_i^{BC}}$ where the $\lambda_i$ correspond to the Schmidt coefficients \cite{schmidt1907,ekert1995}. As such the fidelity is expressed as $\mathcal{F}_{BC} = \sum_i \lambda_i \left| \langle\tilde{\Psi}_i^{BC} | \Phi_{\mathrm{eff}}^{BC}\rangle\right|^2$.}.
The fidelity is provided in Figure \ref{fig:induced_interaction}(c) as a function of $g_{AC}$ and for different fixed values of $g_{AB}$. It becomes apparent that $H^{(2),\mathrm{eff}}$ is not valid for $g_{AA} < g_{A\sigma}$ where the respective impurity phase separates with the bath.
We further note that especially in the regime where the impurities are anti-correlated and share no significant spatial overlap, an effective treatment considering a contact interaction potential fails to describe the full many-body calculations. Instead, in this interaction regime, due to the presence of non-local correlations, a more appropriate choice to model effective impurity-impurity interactions would be a long-range interaction potential, such as the one used in Appendix~\ref{ap:2bfit_Casimir}.
Still, within this effective two-body model different observables for the impurities such as their residue and correlation functions can be extracted  and shown to exhibit a qualitative correct behavior. Deviations from the full many-body results are mostly traced back to the absence of intracomponent  correlations of the bath and impurity-medium ones.

\section{Bipolaron formation}
\label{sec:bip} 

\begin{figure}[t]
\centering
\includegraphics[width=1.0\linewidth]{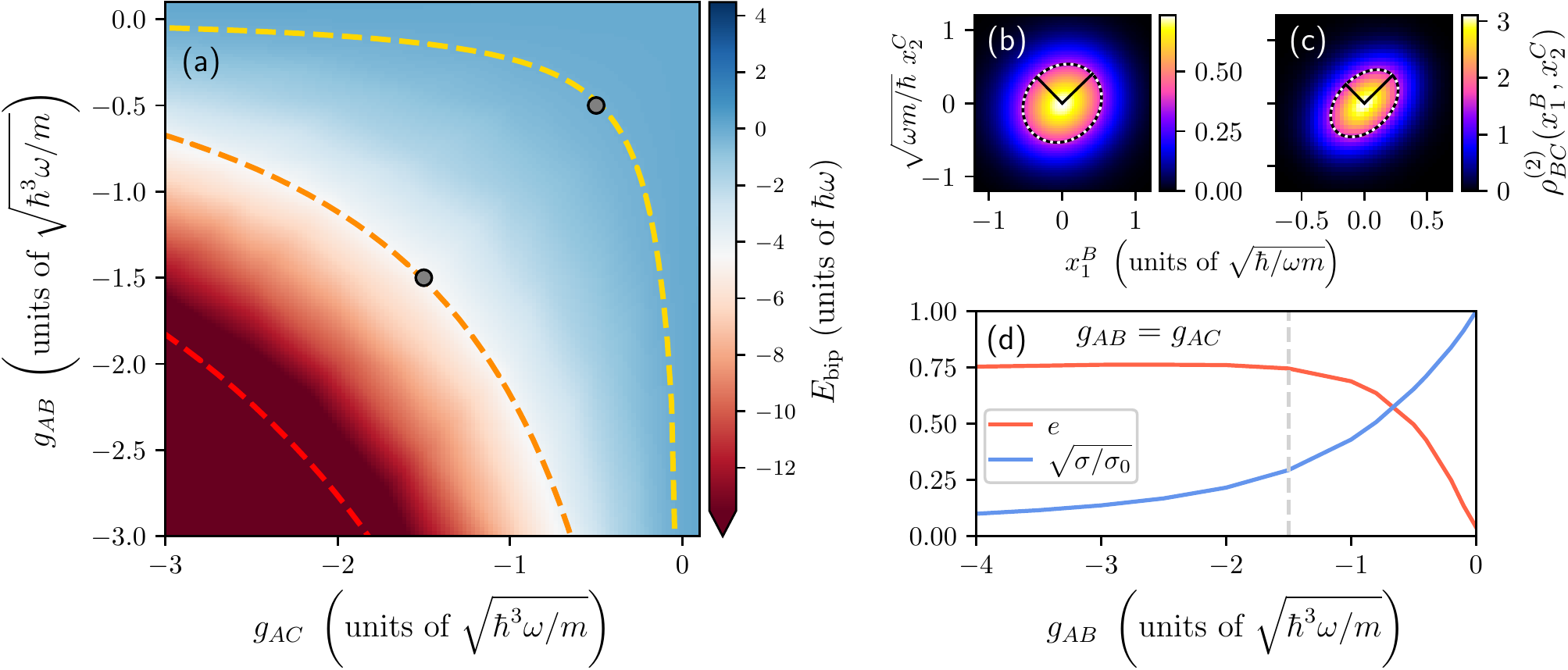}
\caption{(a) Bipolaron energy, $E_{\mathrm{bip}}$, as a function of the intercomponent coupling strengths $g_{AB}$ and $g_{AC}$. The dashed lines represent contours along which the size of the dimer state $\sigma$ remains fixed and in particular from bottom left to top right correspond to $\sqrt{\sigma/\sigma_0} \approx 0.18, 0.29, 0.65$. 
(b), (c) Reduced two-body impurities' density $\rho_{BC}^{(2)}(x_1^B, x_2^{C})$ for $(g_{AB}, g_{AC}) = (-0.5, -0.5)$ and $(-1.5, -1.5)$, respectively, in units of $m\omega/\hbar$ [see also corresponding gray dots in panel (a)]. The region where $\rho_{BC}^{(2)}(x_1^B, x_2^{C})=\rho_{\sigma\sigma'}^{(2)}(0, 0)/2$ is fitted to an ellipse (white dotted line) and shown together with the semi-minor and semi-major axis (black lines). The corresponding eccentricity is depicted in panel (d) assuming $g_{AB}=g_{AC}$. The transition to a bipolaron state where the eccentricity saturates for increasing impurity-medium attractions and the size of the dimer state is $\sqrt{\sigma/\sigma_0}\approx0.29$ occurs at $g_{AC}=-1.5$ (gray dashed line). We consider two non-interacting but distinguishable impurities immersed in a bosonic gas of $N_A = 15$ atoms with $g_{AA} = 0.2$.}
\label{fig:bipolaron}
\end{figure}

Strong attractive induced interactions between two dressed impurities, commonly occurring for strong attractive impurity-medium direct interactions, eventually lead to the formation of a bound dimer quasi-particle state, the so-called bipolaron~\cite{camacho-guardian2018,will2021}. 
In order to probe the presence of such a dimer impurity bound state in our setup, we study the bipolaron energy,
\begin{equation}
    E_{\mathrm{bip}}(g_{AB}, g_{AC}) = E(g_{AB}, g_{AC}) - E_1(g_{AB}) - E_1(g_{AC}) + E_0.
\end{equation}
Here, $E(g_{AB}, g_{AC})$ denotes the total energy of the system including the two distinguishable impurities, $E_0$ is the energy of the bosonic gas in the absence of impurities and $E_1(g_{AB})$, $E_1(g_{AC})$ is the energy of one impurity coupled to the bath. The bipolaron energy is presented in Figure~\ref{fig:bipolaron}(a) covering a wide range of attractive and repulsive impurity-medium interactions, $g_{AB}$ and $g_{AC}$. It features a rapid decrease when both impurities couple attractively to the medium, thereby, evincing the formation of a bound state\footnote{The bipolaron energy decreases exponentially if both impurity-medium couplings ($g_{AB}$, $g_{AC}$) are equally varied from the non-interacting limit to the strongly attractive regime, i.e., along the diagonal in Figure~\ref{fig:bipolaron}(a).}.

A complementary observable used for the identification of the bipolaron is the spatial size of this dimer state. This is naturally captured by $\sigma\sim \sqrt{\langle r_{BC}^2 \rangle}$, where $\langle r_{BC}^2\rangle$ is the squared relative distance [cf. Eq.~(\ref{eq:rel_dist})] between the impurities $B$ and $C$~\cite{camacho-guardian2018}. Specifically, in the following, we track $\sqrt{\sigma/\sigma_0}$ with $\sigma_0$ being the distance in the uncoupled scenario, i.e., at $g_{AB}=g_{AC}=0$, such that we explicitly estimate the impact of the impurity-medium interactions on the dimer size. This is depicted in Figure \ref{fig:bipolaron}(a) as contour dashed lines along which $\sqrt{\sigma/\sigma_0}$ is constant in the $g_{AB}$-$g_{AC}$ plane on top of the bipolaron energy. It can be readily seen that for increasing magnitude of the attractive impurity-medium couplings, i.e., $g_{AB}$ and $g_{AC}$, the size of the dimer state shrinks further, see in particular the dashed lines in Figure~\ref{fig:bipolaron} which from bottom left to top right correspond to $\sqrt{\sigma/\sigma_0} \approx 0.18, 0.29, 0.65$.

The bipolaron dimer state refers to the bunching behavior of the impurities at the same spatial region which manifests in the elongated shape of their two-body density $\rho_{BC}^{(2)}(x_1^B, x_2^{C})$ along the diagonal. In the non-interacting case, i.e., $g_{AB}=g_{AC}=0$, $\rho_{BC}^{(2)}(x_1^B, x_2^{C})$ is circularly symmetric in the $x_1^B-x_2^{C}$ plane and becomes gradually elongated for larger attractions due to the mediated attraction between the impurities, see e.g. Figures \ref{fig:bipolaron}(b) and (c) for the cases $(g_{AB}, g_{AC}) = (-0.5, -0.5)$ and $(-1.5, -1.5)$, respectively, also marked as gray dots in Figure~\ref{fig:bipolaron}(a). To quantify the degree of the aforementioned elongation, we fit the half maximum of the impurities' two-body density\footnote{
We remark that choosing $\rho_{BC}^{(2)}(0, 0)/2$ for the fitting is employed for convenience. Indeed, also other density values were used, e.g. $\rho_{BC}^{(2)}(0, 0)/4$, verifying the same behavior of the eccentricity.
}, i.e. $\rho_{BC}^{(2)}(0, 0)/2$ to a rotated ellipse [see white dotted lines in Figures~\ref{fig:bipolaron}(b) and (c)] and determine the corresponding eccentricity $e=\sqrt{1-b^2/a^2}$ where $a$ ($b$) denotes the semi-major (semi-minor) axis marked by the black lines of the ellipse\footnote{
For the fitting we use the general ellipse equation $\alpha x_1^2 + \beta x_1x_2 + \gamma x_2^2 + \delta x_1 + \epsilon x_2 + \phi = 0$, which in the frame of the ellipse reduces to $\Tilde{x_1}^2/a^2 + \Tilde{x_2}^2/b^2=1$.
}. Apparently for $e=0$, $\rho_{BC}^{(2)}(x_1^B, x_2^{C})$ is circularly symmetric while in the case of $e<1$ it is elongated having the shape of an ellipse. 

The eccentricity of the impurities' two-body density is depicted in Figure \ref{fig:bipolaron}(d) for $g_{AB}=g_{AC}$. By tuning the impurity-medium coupling from the non-interacting limit towards strong attractions, $e$ increases from $e\approx0$ at $g_{AB}=g_{AC}=0$ to finite positive values until it saturates at around $g_{AB}\approx -1.5$. A larger attraction leads only to an additional shrinking of the dimer size, see in particular the exponential decrease of $\sqrt{\sigma/\sigma_0}$ in Figure \ref{fig:bipolaron}(d), leaving the shape of $\rho_{BC}^{(2)}(x_1^B, x_2^{C})$ almost unchanged. In this sense, we deduce that the bipolaron state is formed at $g_{AB}=g_{AC}\approx -1.5$ corresponding to $\sqrt{\sigma/\sigma_0} \approx 0.29$ [vertical gray dashed line in Figure~\ref{fig:bipolaron}(d)]. This observation allows us
to generalize our conclusions for the bipolaron formation also in the case of $g_{AB}\neq g_{AC}$ from the critical size of the dimer state being $\sqrt{\sigma/\sigma_0}\lesssim 0.29$, which corresponds to the central contour dashed line in Figure~\ref{fig:bipolaron}(a).

We remark that the above-described behavior of both $E_{\mathrm{bip}}(g_{AB}, g_{AC})$ and $\sigma/\sigma_0$ is in accordance with previously studied two-component systems containing two indistinguishable bosonic impurities that form a bipolaron\footnote{
We have also verified that upon considering two indistinguishable bosonic impurities our results regarding the bipolaron energy, dimer size and eccentricity coincide with those of the three-component setup with $g_{AB}=g_{AC}$.}
in the strongly attractive coupling regime~\cite{camacho-guardian2018}. However, our results generalize these findings demonstrating the existence of a bipolaron in the case of two distinguishable impurities and suggesting that this bound state is robust to individual variations of $g_{AB}$ or $g_{AC}$ as indicated by the contour lines in Figure~\ref{fig:bipolaron}. Another aspect that we have addressed is that increasing the mass of one impurity, e.g. considering $m_B=2$, leads to a faster reduction of the dimer state size as well as the bipolaron energy for decreasing $g_{AB}=g_{AC}$ while the eccentricity saturates at smaller impurity-medium attractions as compared to the mass-balanced case. This suggests, as expected, that a heavier impurity facilitates bipolaron formation.

\section{Three-body correlations and trimer state}
\label{sec:3bd}

\begin{figure}[t]
\includegraphics[width=1\linewidth]{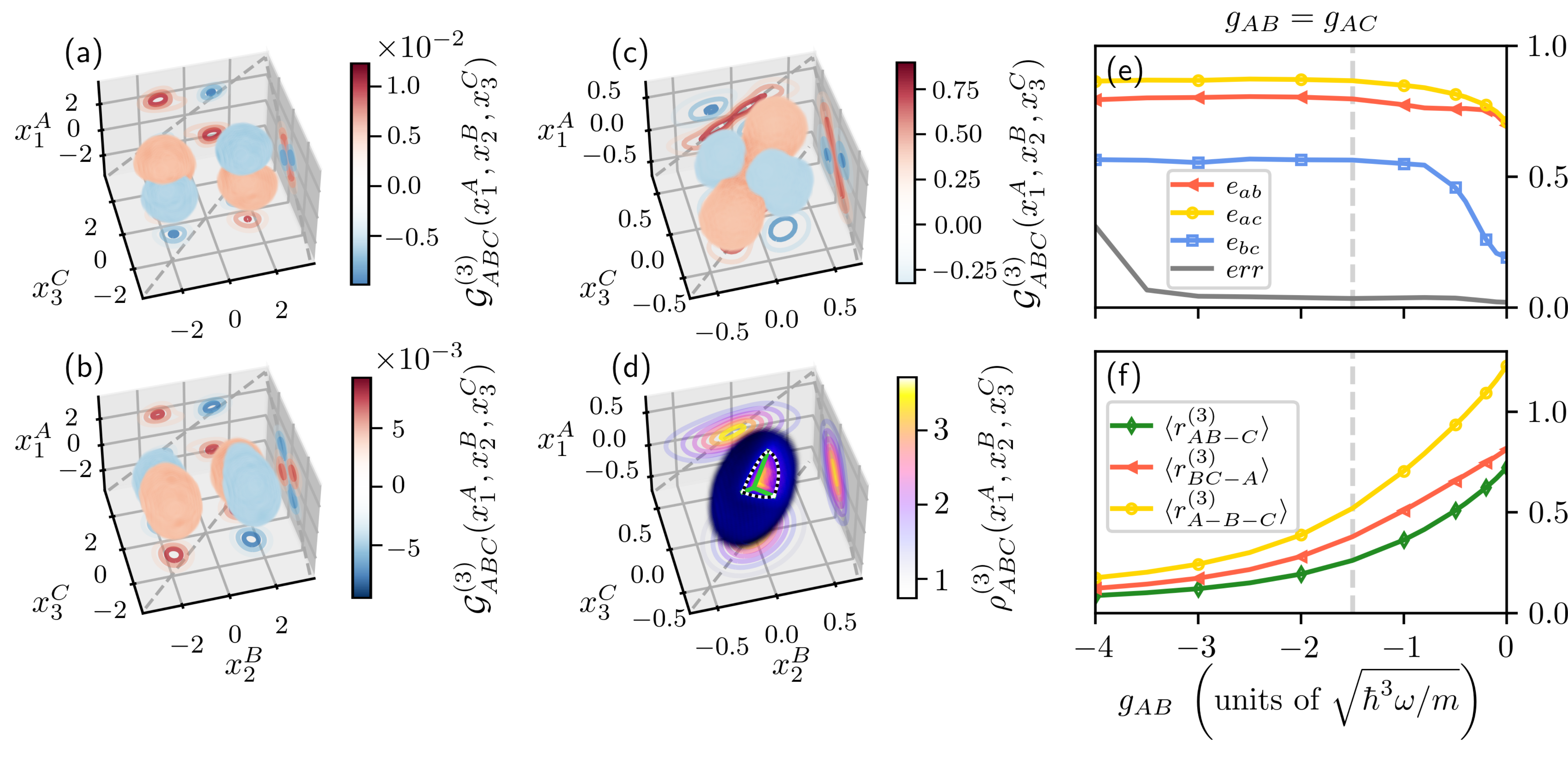}
\caption{(a)-(c) Reduced three-body correlation function $\mathcal{G}_{ABC}^{(3)}(x_1^A, x_2^B, x_3^C)$ for $(g_{AB}, g_{AC}) = (1.0, -0.2)$, $(1.0, 0.2)$ and $(-1.5, -1.5)$, respectively and (d) reduced three-body density $\rho_{ABC}^{(3)}(x_1^A, x_2^B, x_3^C)$ for $(g_{AB}, g_{AC}) = (-1.5, -1.5)$. In each panel, the contours of either (a)-(c) the two-body correlation functions, i.e., $\mathcal{G}_{AB}^{(2)}$, $\mathcal{G}_{AC}^{(2)}$, $\mathcal{G}_{BC}^{(2)}$, or (d) the two-body density functions, i.e., $\rho_{AB}^{(2)}$, $\rho_{AC}^{(2)}$, $\rho_{BC}^{(2)}$, are provided in the $x_1^A-x_2^B$-, $x_1^A-x_3^C$-, $x_2^B-x_3^C$-planes. The spatial coordinates $x^\sigma$ are expressed in units of $\sqrt{\hbar/m\omega}$, whereas $\rho_{ABC}^{(3)}$ and $\mathcal{G}_{ABC}^{(3)}$ are presented in units of $(m\omega/\hbar)^{3/2}$. For visualization purposes we only show the data whose correlation or density value is larger than 0.2 of the respective maximum value. The region corresponding to $\rho_{ABC}^{(3)}(x_1^A, x_2^B, x_3^C)=\rho_{ABC}^{(3)}(0, 0, 0)/2$ is fitted to an ellipsoid rotated in space (part of the fitted ellipsoid is marked by the white dashed lines). The three semi-axis are denoted by the green lines in panel (d). (e) Eccentricities calculated from the semi-axis (see main text) for attractive $g_{AB}=g_{AC}$. (f) Jacobi relative distances $\langle r_{AB-C}^{(3)}\rangle$ and $\langle r_{BC-A}^{(3)}\rangle$ [Eq. (\ref{eq:rel_dist_3_AB_C})] as well as the hyperspherical radius $\langle r_{A-B-C}^{(3)}\rangle$ [Eq. (\ref{eq:rel_dist_3_A_B_C})] for $g_{AB}=g_{AC}$. We mark the transition to a trimer state at $g_{AC}=-1.5$ [gray dashed line in panels (e) and (f)]. For the three-component setup two non-interacting but distinguishable impurities immersed in a bosonic gas of $N_A = 15$ atoms with $g_{AA} = 0.2$ are considered.} 
\label{fig:3drm}
\end{figure}

In the following, we aim to shed light on the existence of three-body correlations appearing in the ground state of the two distinguishable impurities embedded into the bosonic gas. For this purpose, we construct as a first step the normalized reduced three-body density
\begin{align}
    \rho_{ABC}^{(3)}(x_1^A, x_2^B, x_3^C) = \bra{\Psi^{\mathrm{MB}}}
    \hat{\Psi}_A^\dagger(x_1^A) \hat{\Psi}_B^\dagger(x_2^B) \hat{\Psi}_C^\dagger(x_3^C) \hat{\Psi}_C(x_3^C) \hat{\Psi}_B(x_2^B) \hat{\Psi}_A(x_1^A)\ket{\Psi^{\mathrm{MB}}},
    \label{eq:3bd}
\end{align}
which represents the spatially resolved probability of finding at the same time a representative atom of the medium at position $x_1^A$ and the impurities $B$ and $C$ at positions $x_2^B$ and $x_3^C$~\cite{sakmann2008, leveque2021}. Experimentally, the three-body density could be obtained by detecting simultaneously the positions of the three particles of interest, here, the two impurities and one bath atom, and then average over a sample of experimental absorption images \cite{schweigler2017experimental}. Having defined the three-body density, we construct the spatially resolved three-body correlation function as a straightforward extension of the two-body one defined in Eq.~(\ref{eq:noise_correlation}), i.e.,
\begin{align}
    \mathcal{G}_{ABC}^{(3)}(x_1^A, x_2^B, x_3^C) = \rho_{ABC}^{(3)}(x_1^A, x_2^B, x_3^C) - \rho_{A}^{(1)}(x_1^A)\rho_{B}^{(1)}(x_2^B)\rho_{C}^{(1)}(x_3^C).
    \label{eq:3b_corr}
\end{align}
According to this measure, the three participating particles are correlated (anti-correlated) if $\mathcal{G}_{ABC}^{(3)}(x_1^A, x_2^B, x_3^C)>0$ ($\mathcal{G}_{ABC}^{(3)}(x_1^A, x_2^B, x_3^C)$ $<0$), whilst a vanishing $\mathcal{G}_{ABC}^{(3)}(x_1^A, x_2^B, x_3^C)=0$ implies that they are uncorrelated.
Note, that this measure still contains two-body correlation effects since only the product of one-body densities has been subtracted from the three-body density.

The three-body correlation function is depicted in Figures~\ref{fig:3drm}(a) and (b) for the case of strong repulsions between impurity $B$ and the bath ($g_{AB}=1$) and either weak attractive or repulsive couplings between the bath and the $C$ impurity, namely $g_{AC}=-0.2$ and $0.2$, respectively. Moreover, for visualization and completeness issues, we additionally showcase within the $x_1^A$-$x_2^B$, $x_1^A$-$x_3^C$ and $x_2^B$-$x_3^C$ planes the underlying two-body correlation functions $\mathcal{G}_{AB}^{(2)}(x_1^A, x_2^B)$, $\mathcal{G}_{AC}^{(2)}(x_1^A, x_3^C)$ and $\mathcal{G}_{BC}^{(2)}(x_2^B, x_3^C)$, respectively\footnote{
As an example, notice that the contours in the $x_1^A$-$x_2^B$ and $x_2^B$-$x_3^C$ planes of Figure~\ref{fig:3drm}(c) correspond to the $\mathcal{G}_{AB}^{(2)}(x_1^A, x_2^B)$ and $\mathcal{G}_{AB}^{(2)}(x_1^B, x_2^C)$ illustrated in Figures~\ref{fig:corr}(b1) and (b2), respectively.}. 
Focusing on $g_{AC}=-0.2$, it becomes evident that $\mathcal{G}_{ABC}^{(3)}(x_1^A, x_2^B, x_3^C)$ fragments into two correlated and two anti-correlated parts. The correlated segments indicate that it is likely for one bath atom and the $C$ impurity to reside at the same side with respect to the trap center while the repulsively coupled impurity $B$ favors to be on the opposite side. On the other hand, the anti-correlated fragments suggest that a configuration where the impurities and a bath atom are at the same location is not favorable. The spatial arrangement of these fragments is altered in the three-dimensional space if the sign of $g_{AC}$ is inverted, in a sense that the correlated and anti-correlated regions are rotated by roughly 90$^{\circ}$ around the $x_2^B$ direction. 
In such a configuration the impurities are located at the same side in terms of the trap center and a bath atom lies on the opposite side. The corresponding two-body correlation functions $\mathcal{G}_{AC}^{(2)}(x_1^A, x_3^C)$ and $\mathcal{G}_{BC}^{(2)}(x_2^B, x_3^C)$ become inverted, whereas $\mathcal{G}_{AB}^{(2)}(x_1^A, x_2^B)$ preserves its pattern, see the contours in Figures \ref{fig:3drm}(a) and (b). 

Subsequently, we turn to strongly attractive impurity-medium interactions with $g_{AB}=g_{AC}$. Here, the three-body density $\rho_{ABC}^{(3)}(x_1^A, x_2^B, x_3^C)$ becomes elongated exhibiting an ellipsoidal shape, see e.g. Figure~\ref{fig:3drm}(d) for $(g_{AB},g_{AC})=(-1.5, -1.5)$. Thereby, the three-body density is stretched along the $(x_1^A, x_2^B, x_3^C)$-direction, i.e., the diagonal of the coordinate system, demonstrating a bunching behavior of the two impurities and a representative atom of the bath species.
In particular, the corresponding three-body correlation function, presented in Figure~\ref{fig:3drm}(c), features a correlated pattern along the diagonal around which a shell-like structure consisting of anti-correlated fragments is formed. 

To quantify the deformation of the three-body density, we fit its half maximum, i.e., $\rho_{ABC}^{(3)}(0, $ $0, 0)/2$, to a rotated ellipsoid (see white dashed lines in Figure~\ref{fig:3drm}(d) corresponding to a profile of the ellipsoid). Specifically, we fit the ellipsoid equation $\Tilde{x_1}^2/a^2 + \Tilde{x_2}^2/b^2 + \Tilde{x_3}^2/c^2=1$, where $\Tilde{x_i}$ refers to the coordinate system of the ellipsoid spanned by its semi-axis with lengths $a$, $b$ and $c$ [green lines in Figure \ref{fig:3drm}(d)]. From the semi-axis we determine three eccentricities, namely $e_{ab}=\sqrt{1-b^2/a^2}$, $e_{ac}=\sqrt{1-c^2/a^2}$ and $e_{bc}=\sqrt{1-c^2/b^2}$ with $a\geq b\geq c$. These eccentricities are depicted in Figure~\ref{fig:3drm}(e) together with the relative deviation, $err$, from the ellipsoid function for varying $g_{AB}$ and assuming $g_{AB}=g_{AC}$. In the non-interacting case, i.e., $g_{AB}=g_{AC}=0$, the eccentricities are already finite indicating a deviation from a spherical shape, which is in contrast to the bipolaron [cf. Figure \ref{fig:bipolaron}(d)]. 
This is attributed to the presence of finite intraspecies interactions among the bath particles causing the observed spatial deformation. Importantly, the eccentricities show an increasing tendency for stronger attractive values of $g_{AB}=g_{AC}$, meaning that the elongation of the ellipsoid is enhanced until it saturates at around $g_{AB}=g_{AC}\approx-1.5$. 

A further characterization of the size of the three-body cluster at strong attractions is achieved by inspecting the hyperspherical radius $\langle r_{\sigma - \sigma' - \sigma''}^{(3)} \rangle$ and the Jacobi relative distance $\langle r_{\sigma' \sigma'' - \sigma}^{(3)} \rangle$. The latter denotes the distance between the atom $\sigma$ and the center-of-mass of the particles $\sigma'$ and $\sigma''$ \cite{whitten1968, greene2017, bougas2021a}. These observables are defined as 
\begin{align}
    \langle r_{A - B - C}^{(3)} \rangle = \frac{1}{N_A N_{B} N_{C}}\int {\mathrm{d}} x_1^A {\mathrm{d}} x_2^B {\mathrm{d}} x_3^C \sqrt{ (x_1^A)^2 + (x_2^B)^2 + (x_3^C)^2 } \rho_{ABC}^{(3)}(x_1^A, x_2^B, x_3^C)
    \label{eq:rel_dist_3_A_B_C}, 
    \\
    \langle r_{\sigma' \sigma'' - \sigma}^{(3)} \rangle = \frac{1}{N_A N_{B} N_{C}}\int {\mathrm{d}} x_1^A {\mathrm{d}} x_2^B {\mathrm{d}} x_3^C \left|x^\sigma - \frac{1}{2}\left( x^{\sigma'} + x^{\sigma''} \right) \right| \rho_{ABC}^{(3)}(x_1^A, x_2^B, x_3^C),
    \label{eq:rel_dist_3_AB_C}
\end{align}
with $\sigma, \sigma', \sigma'' \in \{A,B,C\}$ and $\sigma\neq\sigma'$, $\sigma\neq\sigma''$, $\sigma'\neq\sigma''$. Note that in the present case $\langle r_{AB - C}^{(3)} \rangle = \langle r_{AC - B}^{(3)} \rangle$, since impurity $B$ and $C$ have identical mass and are coupled with the same strength to the bath. Figure \ref{fig:3drm}(f) reveals that for stronger impurity-medium attractions the hyperspherical radius decreases exponentially implying an exponential shrinking of the size of the three-body cluster. The same exponential decrease is also captured by the expectation values of the Jacobi relative distances where we find $\langle r_{BC - A}^{(3)} \rangle < \langle r_{AB - C}^{(3)} \rangle$ reflecting the fact that the bath atoms extend over a larger spatial region than the impurities due to the repulsive $g_{AA}$. 
The above properties imply the formation of a bound trimer state for couplings $g_{AB}=g_{AC}\leq -1.5$ corresponding to values where the ellipsoidal structure of the three-body density saturates. In this sense, the formation of a bipolaron is accompanied by the development of a bound trimer state.

\section{Conclusions and perspectives}
\label{sec:conclusion}

We have studied the correlation properties in the ground state of 
two non-interacting distinguishable impurities immersed in a bosonic bath with the entire three-component system being harmonically trapped. The impurities become dressed by the excitations of the bosonic gas generating quasiparticle states, herein Bose polarons, having characteristic properties such as effective mass and featuring induced correlations. In order to appreciate the impact of inter- and intracomponent correlations we rely on the variational ML-MCTDHX method whose flexible wave function truncation ansatz allows to operate at different correlation orders. An emphasis is placed on the high tunability of the three-component setting unveiling rich density and correlation patterns, the manipulation of both the sign and the strength of impurities induced interactions as well as the formation of bound impurity states. 

Specifically, we demonstrate that upon varying the involved impurity-medium couplings, both impurities can either localize at the trap center (attractive intercomponent interactions), form a shell around the bosonic gas (repulsive interactions), i.e., phase-separate, or one of them localize and the other phase-separate (alternating signs of impurity-medium couplings). These density configurations can be understood at least qualitatively in terms of an effective potential picture for the impurities which refers to a dipped harmonic oscillator (double-well) for attractive (repulsive) intercomponent interactions. 

A detailed characterization of the induced correlations is provided in a wide range of impurity-medium interactions aiming to expose their intricate role. Inspecting the two-body intercomponent correlation functions we find that the bosonic gas mediates anti-correlations among the impurities if one of them couples repulsively and the other attractively to it. In contrast, induced two-body correlations occur as long as both impurities couple either attractively or repulsively to their medium. The origin of the aforementioned correlation patterns is traced back to the spatial configurations of each component. This means that if the impurities have a finite spatial overlap with the bath the latter mediates two-body correlations between them. Interestingly, there is also the possibility that the impurities are not overlapping but can be still correlated implying that non-local correlations are in play. To quantify the strength and sign of the induced interactions we employ the relative two-body distance among the impurities extracting all contributions stemming from mean-field effects. 
In this sense, it is demonstrated that induced two-body correlations (anti-correlations) are related to mediated attractive (repulsive) impurity interactions. 
These findings are further supported by an effective two-body model containing the impurities effective trapping potential and their induced interactions. Importantly, this approach allows to determine the strength and sign of the effective interactions mediated between the impurities through a comparison with the full many-body results. Moreover, by constructing an effective one-body Hamiltonian enables us to estimate the effective mass and trapping frequency of each distinguishable impurity (polaron), see Appendix~\ref{ap:1dfit}.

Evidences regarding bipolaron formation are provided, when both impurities are strongly attractively coupled to the bosonic gas, by means that the bipolaron energy and the size of the underlying dimer state rapidly decrease for stronger attraction. Interestingly, we determine the intercomponent three-body correlation function according to which overall weak three-body correlations exist and become enhanced for strongly attractive impurity-medium interactions signaling the formation of trimers among the impurities and an atom of the medium.

In this investigation we have restricted ourselves to the ground state of the three-component mixture. Further understanding on the character of the impurities induced interactions and in particular their nonlocal character and their dependence on the statistics of the medium are interesting perspectives. 
In this context, a systematic finite size scaling analysis with respect to the number of bath particles in order to infer  the persistence of our findings e.g. in terms of the crossover of the impurities induced interactions (see also Appendix~\ref{ap:mB2_NA30}) and in general the build-up of intercomponent correlations would be desirable as well.
Also, the emulation of spectroscopic schemes that will allow the identification of the ensuing polaron states and excitations \cite{jorgensen2016, mistakidis2020} constitutes an intriguing direction.
Furthermore, studying the behavior of impurities induced interactions and bound states in different external trapping potentials is also an interesting direction. 
%intercomponent correlation behavior for different trapping potentials might be an interesting future project. 
Here, a setup of immediate interest would be to load the bath atoms in a ring potential and investigate the formation of impurities bound states in both the attractive and the repulsive impurities-medium interaction regimes.
Another straightforward extension would be to explore the nonequilibrium impurities dynamics in order to understand the build-up of induced correlations. 
An additional fruitful research direction is to understand the Bose polaron formation when indistinguishable impurities are immersed in an attractive two-component gas forming a droplet. Certainly, studying correlation effects in particle-balanced three component settings with an emphasis on the few- to many-body crossover and in particular close to the pair immiscibility threshold is worth to be pursued.

\section*{Acknowledgements}
This work has been funded by the Deutsche Forschungsgemeinschaft (DFG, Germany Research Foundation) --- SFB 925 --- project 170620586. S.I.M. gratefully acknowledges financial support from the NSF through a grant for ITAMP at Harvard University.

\begin{appendix}

\section{Behavior of the bipartite entanglement}
\label{ap:entanglement}

A standard measure to estimate the bipartite entanglement of mixed states that exist in a multi-component system\footnote{Notice that, for instance, the von-Neumann entropy as an entanglement measure is well-defined in a two species but it is not applicable in multi-component ones~\cite{horodecki2009}.} is encapsulated in the logarithmic negativity~\cite{vidal2002,horhammer2008,duarte2009,zell2009,shiokawa2009,charalambous2019,becker2022}. It is based on the partial transpose of the two-body species reduced density matrix\footnote{
This is completely different from the two-body density matrix of two particles given by Eq.~(\ref{eq:2bd}).}
, which, e.g. referring to species $\sigma$ and $\sigma'$, is obtained by integrating out the degrees of freedom of species $\sigma''$ leading to $\rho_{\sigma\sigma'}^{(2),\mathrm{spec}}=\mathrm{Tr}_{\sigma''}\left(|\Psi^{\mathrm{MB}}\rangle \langle\Psi^{\mathrm{MB}}| \right) = \sum_{ijlm}\sum_kC_{ijk}C_{lmk}^* |\Psi_i^{\sigma}\rangle |\Psi_j^{\sigma'}\rangle\langle\Psi_l^{\sigma}|\langle\Psi_m^{\sigma'}|$ \cite{zyczkowski1998,braun2002, benatti2003}.

Its partial transpose $T_\sigma$ with respect to species $\sigma$ is calculated by exchanging the indices $i$ and $l$ associated with species $\sigma$, i.e., $\left(\rho_{\sigma\sigma'}^{(2),\mathrm{spec}}|_{ijlm}\right)^{T_\sigma}=\rho_{\sigma\sigma'}^{(2),\mathrm{spec}}|_{ljim}$. Calculating the eigenvalues of $\left(\rho_{\sigma\sigma'}^{(2),\mathrm{spec}}\right)^{T_\sigma}$ and in particular summing up its negative eigenvalues $\mu_i$ yields the so-called negativity, $\mathcal{N}_{\sigma\sigma'}=\sum_i|\mu_i|$. Subsequently, the logarithmic negativity reads
\begin{align}
    \mathcal{E}_{\sigma\sigma'} = \log_2\left(1+2\mathcal{N_{\sigma\sigma'}}\right).
\label{eq:log_neg}
\end{align}
This measure exploits the fact that for a separable mixture, e.g. $\rho_{\sigma\sigma'}^{(2),\mathrm{spec}} = \sum_i p_i \tilde{\rho}_{\sigma, i}^{(1),\mathrm{spec}}\otimes\tilde{\rho}_{\sigma',i}^{(1),\mathrm{spec}}$, the partial transpose does not alter the spectrum of $\rho_{\sigma\sigma'}^{(2),\mathrm{spec}}$ and, hence, all eigenvalues remain positive. In this sense, the presence of negative eigenvalues guarantees the existence of entanglement. However, this statement can not be inverted, i.e., even if the logarithmic negativity is zero the species $\sigma$ and $\sigma'$ can still be entangled~\cite{horodecki2009}.

\begin{figure*}[t]
    \centering
    \includegraphics[width=0.65\linewidth]{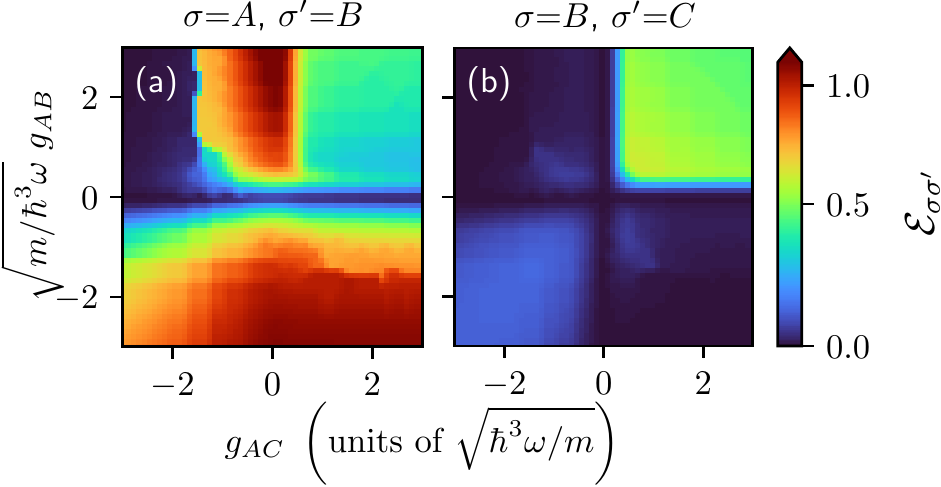}
    \caption{(a)-(b) Diagram of the intercomponent (see legends) logarithmic negativity $\mathcal{E}_{\sigma\sigma'}$ [Eq.~(\ref{eq:log_neg})] as a function of the impurity-medium couplings ($g_{AB}$, $g_{AC}$). The harmonically trapped three component system consists of two non-interacting but distinguishable impurities immersed in a bosonic gas of $N=15$ atoms with $g_{AA}=0.2$.}
    \label{fig:phase_diag_ent}
\end{figure*}

The logarithmic negativity between the bath and the $B$ impurity, $\mathcal{E}_{AB}$, as well as among the impurities, $\mathcal{E}_{BC}$, is illustrated in Figures~\ref{fig:phase_diag_ent}(a) and (b) respectively within the $g_{AB}$-$g_{AC}$ plane. As expected it overall captures the main features of the integrated correlation functions shown in Figures~\ref{fig:phase_diag}(a) and (b). For instance, $\mathcal{E}_{AB}$ vanishes for strongly attractive $g_{AC}$ and strongly repulsive $g_{AB}$ [Figure \ref{fig:phase_diag_ent}(a)], while the parameter region referring to the impurities coalescence is in a similar way pronounced in $\mathcal{E}_{BC}$ as it has been observed for $\mathcal{C}_{BC}$, compare Figures~\ref{fig:phase_diag}(a) and (b) for repulsive $g_{AB}$ and $g_{AC}$. Recall that while $\mathcal{E}_{\sigma \sigma'}$ provides only a quantitative diagnostic for the bipartite entanglement and does not describe the correlated or anti-correlated behavior as $\mathcal{C}_{\sigma \sigma'}$ it still gives insight into the entanglement content of the many-body system. As such, for large $g_{AB}<0$ the logarithmic negativity uncovers that the bath and the $B$ impurity are strongly entangled especially so in the repulsive $g_{AC}>0$ region, while varying $g_{AB}$ towards the weakly attractive regime and for $\abs{g_{AC}}>1$ entanglement is reduced [Figure \ref{fig:phase_diag_ent}(a)]. This is attributed to the simultaneous increase of $\mathcal{E}_{AC}$\footnote{
Since the impurities are in this case physically identical, i.e., $m_A=m_B \equiv m$ and $\omega_A=\omega_B \equiv \omega$, the phase diagram of $\mathcal{E}_{AC}$ corresponds to the one of $\mathcal{E}_{AB}$ but reflected along the diagonal $g_{AB}=g_{AC}$.}, unveiling a competition between the intercomponent entanglement of individual impurities with the medium. Finally, in line with the predictions of $\mathcal{C}_{BC}$, $\mathcal{E}_{BC}$ demonstrates that entanglement is finite when both impurities are either weakly attractive or strongly repulsively coupled to the medium, see Figure \ref{fig:phase_diag}(d).

\section{Effective mass and trap frequency of a single impurity}
\label{ap:1dfit}

\begin{figure}[t]
    \centering
    \includegraphics[width=0.8\linewidth]{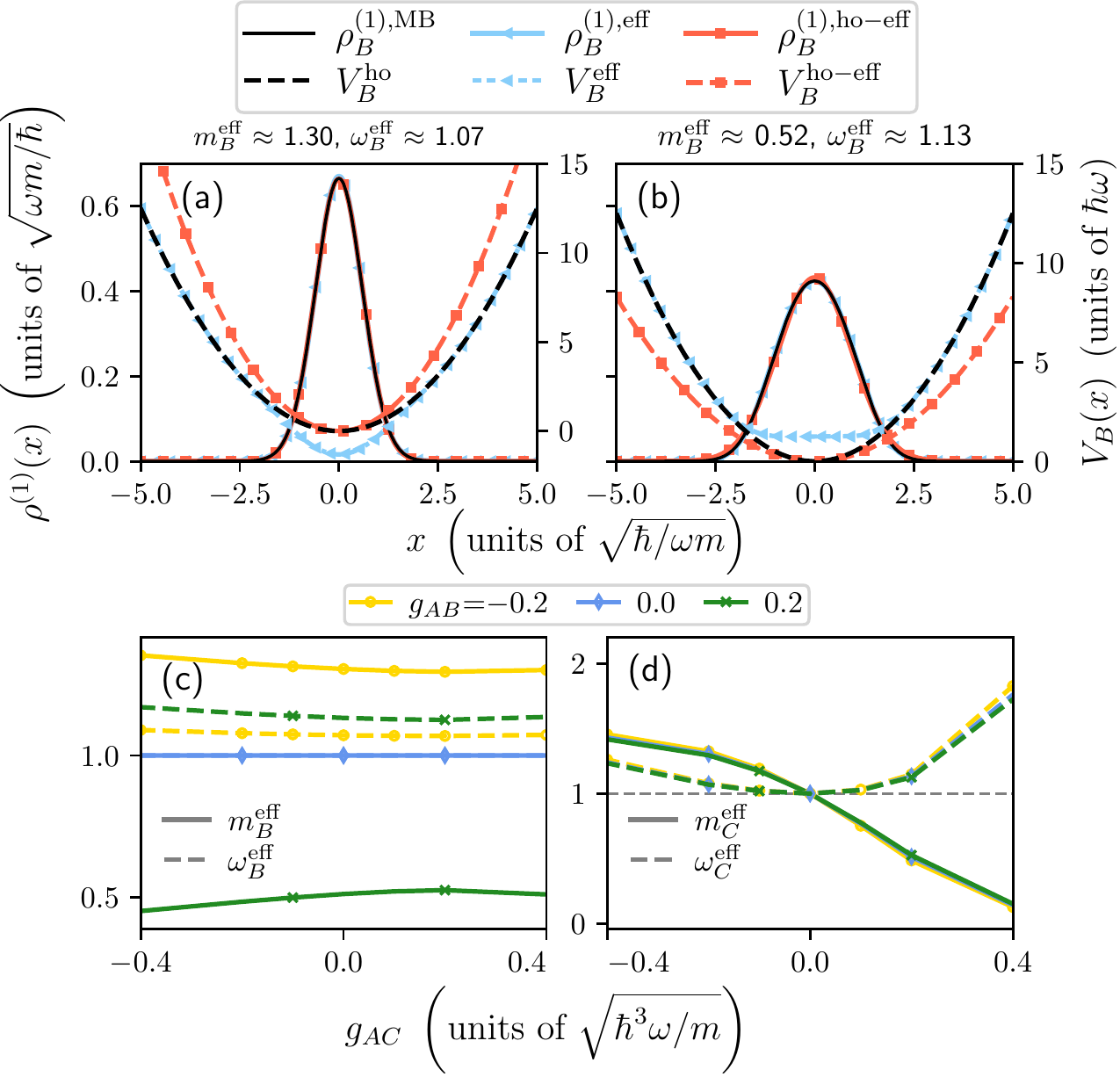}
    \caption{One-body density of the $B$ impurity obtained within different approaches (see legend) for the interaction configurations (a) $(g_{AB}, g_{AC}) = (-0.2, 0.1)$ and (b) $(0.2, 0.1)$. Specifically, $\rho^{(1)}_{\mathrm{MB}}$ denotes the one-body distribution of the full three-component many-body system, whereas $\rho^{(1), \mathrm{eff}}_B$ and $\rho^{(1), \mathrm{ho-eff}}_B$ are calculated using the effective one-body Hamiltonians composed of either an effective harmonic oscillator with an effective mass and frequency [cf. Eq.~(\ref{eq:1body_eff_ham})] or the effective potential defined in Eq.~(\ref{eq:effpot}), respectively. Effective mass and trapping frequency of the dressed (b) $B$ and (c) $C$ impurity, respectively, as deduced from the effective polaron model defined of Eq.~(\ref{eq:1body_eff_ham}).}
    \label{fig:eff_mass}
\end{figure}

In the following, we approach the three-component impurity setting as a polaron problem since each individual impurity via its coupling to the bosonic gas is dressed by the excitations of the latter. In this sense, we aim to capture the effective behavior of the $B$ and $C$ impurity with the effective one-body model~\cite{mistakidis2019}, 
\begin{align}
    \hat{H}_\sigma^{(1), \mathrm{ho-eff}} = - \frac{\hbar^2}{2m_\sigma^{\mathrm{eff}}} \frac{\partial^2}{(\partial x^{\sigma})^2} + \frac{1}{2}m_\sigma^{\mathrm{eff}}(\omega_\sigma^{\mathrm{eff}})^2x^2,
    \label{eq:1body_eff_ham}
\end{align}
where $m_\sigma^{\mathrm{eff}}$ and $\omega_\sigma^{\mathrm{eff}}$ denote the polaron effective mass and trapping frequency with $\sigma\in\{B,C\}$\footnote{
Recall that within the effective two-body model described by Eq.~(\ref{eq:eff_two_body_model}) we implicitly account for the effective mass and frequency via the effective potential $V_{B,C}^{\mathrm{eff}}$ Eq.~(\ref{eq:effpot})].
Indeed, beyond mean-field corrections imprinted on $\rho_A^{(1)}$ and, thus appearing in $V_{B,C}^{\mathrm{eff}}$, affect the effective mass and frequency~\cite{mistakidis2019}.}.
To identify the values of the effective mass and frequency, we minimize the cost function 
\begin{align}
    \mathcal{L}_{\sigma} = \Delta \rho^{(1)}_\sigma + \Delta E_\sigma. 
    \label{eq:1body_cost_fct}
\end{align}
In this expression, the first term refers to $\Delta \rho^{(1)}_\sigma = \int \mathrm{d}x_\sigma \left| \rho^{(1), \mathrm{MB}}_\sigma(x_\sigma) - \rho^{(1), \mathrm{ho-eff}}_\sigma(x_\sigma) \right|^2$ with $\rho^{(1), \mathrm{MB}}_\sigma$ and $\rho^{(1), \mathrm{ho-eff}}_\sigma$ being the one-body density as predicted from the full three-component system and the effective one-body model, respectively. 
The second contribution of the right-hand side in Eq.~(\ref{eq:1body_cost_fct}) designates the energy difference $\Delta E_\sigma = \left|E_\sigma^{\mathrm{MB}} - E_\sigma^{\mathrm{ho-eff}} \right|^2$, where $E_\sigma^{\mathrm{MB}} = $ $ \langle \Psi^{\mathrm{MB}} | \hat{H}_\sigma | \Psi^{\mathrm{MB}} \rangle$ is the $\sigma$ impurity energy and $E_\sigma^{\mathrm{ho-eff}} = \langle \phi | \hat{H}_\sigma^{(1), \mathrm{ho-eff}} | \phi\rangle = \frac{1}{2} \omega_\sigma^{\mathrm{eff}}$ is the energy of the effective one-body model and $|\phi\rangle$ the corresponding ground state. 
Note that in order to uniquely estimate $m_\sigma^{\mathrm{eff}}$ and $\omega_\sigma^{\mathrm{eff}}$ one needs to adequately describe both the density and the energy of the impurity.

Figures~\ref{fig:eff_mass}(a) and (b) showcase the one-body densities $\rho^{(1), \mathrm{MB}}_B$ and $\rho^{(1) \mathrm{ho-eff}}_B$ for the characteristic interaction configurations $(g_{AB}, g_{AC}) = (-0.2, 0.1)$ and $(0.2, 0.1)$, respectively.
For comparison we additionally provide the one-body density $\rho^{(1), \mathrm{eff}}_B$ obtained from $\hat{H}_B^{(1), \mathrm{eff}} = - \frac{\hbar^2}{2m_B} \frac{\partial^2}{(\partial x^B)^2}$ $ + V_B^{\mathrm{eff}}$. 
As it can be readily seen, the one-body densities predicted by the two effective one-body models are in excellent agreement with the one corresponding to the full three-component many-body system. Deviations start to become evident for strong repulsive impurity-medium couplings (not shown) where the impurity and the medium phase separate~\cite{mistakidis2019,mistakidis2022cold}. Recall that the effective model is by definition valid for weak intercomponent repulsions where the impurity does not probe the edges of the bosonic cloud.

The effective masses and frequencies of the $B$ and $C$ impurities after minimization of the cost function given by Eq.~(\ref{eq:1body_cost_fct}) are represented in Figures~\ref{fig:eff_mass}(c) and (d) with respect to the impurity-medium couplings. 
It is important to point out that both the effective mass and frequency of a specific
impurity, e.g. the $B$ one, primarily depend on its coupling with the bath $g_{AB}$. The interaction strength of the other impurity ($C$) with the bath, e.g. $g_{AC}$, has almost no impact on the effective parameters of impurity $B$. 
For instance, this conclusion can be drawn from the nearly constant behavior of $m_B^{\mathrm{eff}}$ and $\omega_B^{\mathrm{eff}}$ for varying $g_{AC}$ shown in Figure~\ref{fig:eff_mass}(c), or the fact that $m_C^{\mathrm{eff}}$ and $\omega_C^{\mathrm{eff}}$ remain almost intact for fixed $g_{AC}$ and different $g_{AB}$, see Figure~\ref{fig:eff_mass}(d).

For an attractively coupled impurity with the bosonic gas, the effective mass and frequency become larger than their bare values [gray dashed lines in Figures~\ref{fig:eff_mass}(c) and (d)], see in particular $m_B^{\mathrm{eff}}$, $\omega_B^{\mathrm{eff}}$ when $g_{AB}=-0.2$ in Figure~\ref{fig:eff_mass}(c) and $m_C^{\mathrm{eff}}$, $\omega_C^{\mathrm{eff}}$ for $g_{AC}<0$ in Figure~\ref{fig:eff_mass}(d). As such, the emergent Bose polaron experiences a narrower trapping potential, thereby, reflecting the localization of the impurity at the trap center [cf. $\rho_B^{(1), \mathrm{ho-eff}}$ and $V_B^{\mathrm{ho-eff}}$ in Figure~\ref{fig:eff_mass}(a)]. On the other hand, in the case of a repulsively coupled impurity the effective trapping frequency is still tighter than the original value, but the effective mass becomes smaller than its bare value [cf. $m_B^{\mathrm{eff}}$, $\omega_B^{\mathrm{eff}}$ for $g_{AB}=0.2$ in Figure~\ref{fig:eff_mass}(c) as well as $m_C^{\mathrm{eff}}$, $\omega_C^{\mathrm{eff}}$ for $g_{AC}>0$ in Figure~\ref{fig:eff_mass}(d)]. In particular, the effective mass is small enough to compensate the increased effective frequency meaning that the underlying harmonic trap is eventually broadened [cf. $V_B^{\mathrm{ho-eff}}$ in Figure~\ref{fig:eff_mass}(b)]. Additionally, the comparatively smaller effective mass is related to a spatial delocalization of the impurity cloud\footnote{
Indeed, the kinetic energy of, e.g., the impurity $C$ increases for increasing $g_{AC}$ while the potential energy remains nearly constant.}.
In this way, the effective one-body model captures the effects imprinted on the impurity in the three-component system.

\section{Modelling the effective impurity interactions with an exponential potential}
\label{ap:2bfit_Casimir}

\begin{figure}[t]
    \centering
    \includegraphics[width=0.85\linewidth]{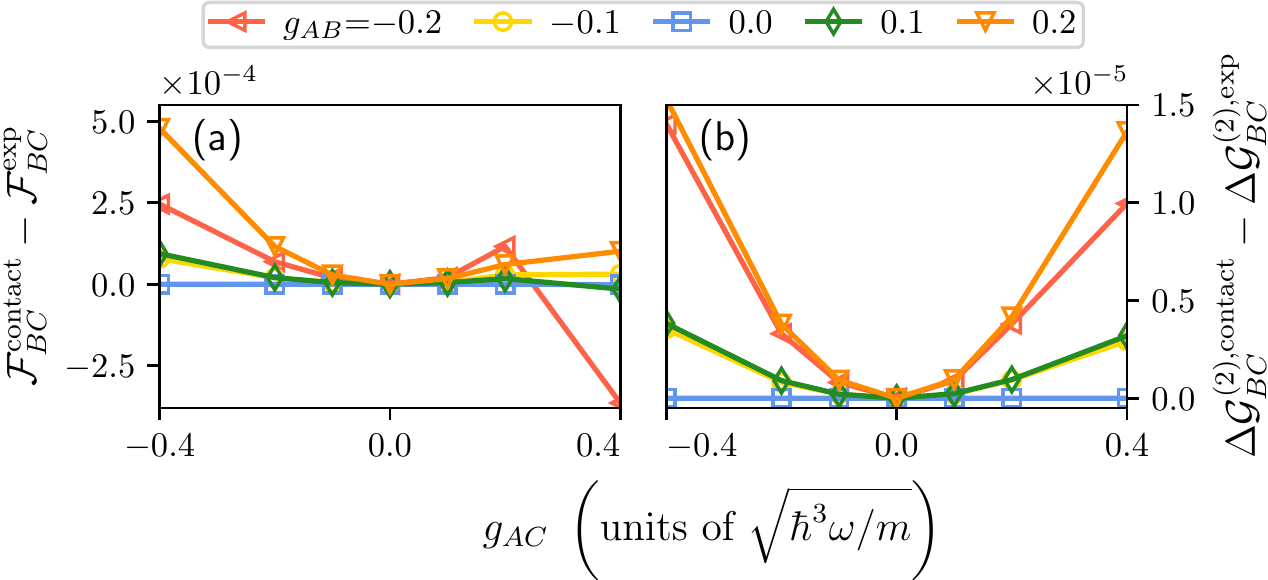}
    \caption{(a) Relative deviation between the fidelities $\mathcal{F}_{BC}^{\mathrm{exp}}$ and $\mathcal{F}_{BC}^{\mathrm{contact}}$, which correspond to the overlap of the impurities two-body wave function obtained from the full many-body approach and the effective model of Eq.~(\ref{eq:eff_two_body_model}) containing either an exponential or a contact-type interaction potential, respectively. (b) Difference between $\Delta \mathcal{G}_{BC}^{(2), \mathrm{exp}}$ and $\Delta \mathcal{G}_{BC}^{(2), \mathrm{contact}}$, referring to the variance of the two-body correlation function calculated within the effective two-body model using either the contact or the exponential interaction potential with respect to the full three-component system. For both quantities the relative deviations are minor, testifying the validity of both effective interaction potentials.}
\label{fig:ind_int_casimir}
\end{figure} 

To verify the validity of the contact interaction potential for describing the induced impurity interactions between the impurities [Eq. (\ref{eq:eff_two_body_model})], we next exemplify that our results do not change if one instead uses an exponential potential. The latter has been derived in Refs.~\cite{reichert2019, reichert2019a} and holds in the homogeneous case and for immobile impurities residing at distances satisfying $l=|x^B-x^C| \ll \xi_A$, with $\xi_A \approx 1/\sqrt{2m_Ag_{AA}N_A\rho_A^{(1)}(0)} \approx 0.6$ being the healing length of the bath. In particular, we replace the interaction term in Eq.~(\ref{eq:eff_two_body_model}) with
\begin{align}
    U(l) = -\frac{g_{AB}g_{AC}m_A}{\sqrt{\gamma}}e^{-2l/\xi_A},
    \label{eq:ind_int_casimir}
\end{align}
where $\gamma=\frac{m_Ag_{AA}}{N_A\rho_A^{(1)}(0)}$\footnote{
We model the exponential potential with the so-called POTFIT method~\cite{jackle1996,jackle1998}.}.
As discussed in Section~\ref{sec:induced_interactions_eff_2b_model}, we judge the quality of the effective two-body model by estimating the fidelity, $\mathcal{F}_{BC}$, between the impurities two-body wave function as extracted from the full many-body system and the effective two-body model containing either a contact or an exponential interaction potential. Subsequently, we determine the difference $\mathcal{F}_{BC}^{\mathrm{exp}} - \mathcal{F}_{BC}^{\mathrm{contact}}$ which as shown in Figure~\ref{fig:ind_int_casimir}(a) testifies deviations at most of the order $10^{-4}$. 

Proceeding one step further, we determine the overlap between the respective two-body correlation functions of the impurities determined within the full three-component system and the effective two-body model. Namely, we track $\Delta \mathcal{G}_{BC}^{(2), \mathrm{exp}} = \int \mathrm{d}x_B\mathrm{d}x_C \left| \mathcal{G}_{BC}^{(2)} - \mathcal{G}_{BC}^{(2), \mathrm{exp}} \right|^2$, where $\mathcal{G}_{BC}^{(2), \mathrm{exp}}$ denotes the two-body correlation function obtained within the effective two-body model (see also Section~\ref{sec:induced_interactions_eff_2b_model}) with an exponential interaction. 
To infer the deviations among the exponential and contact effective interactions at the two-body correlation level, we calculate the difference $\Delta \mathcal{G}_{BC}^{(2), \mathrm{exp}} - \Delta \mathcal{G}_{BC}^{(2), \mathrm{contact}}$, see Figure \ref{fig:ind_int_casimir}(b). Also here, only small deviations of the order $10^{-5}$ are identified. 

Therefore, the contact and exponential effective interaction potentials lead essentially to the same description regarding the impurities properties. This outcome was not {\it a-priori} expected since the exponential potential is originally derived in the homogeneous case.

\section{Impact of mass-imbalanced impurities and the atom number of the bosonic gas} \label{ap:mB2_NA30} 

\begin{figure}[t]
    \centering
    \includegraphics[width=0.65\linewidth]{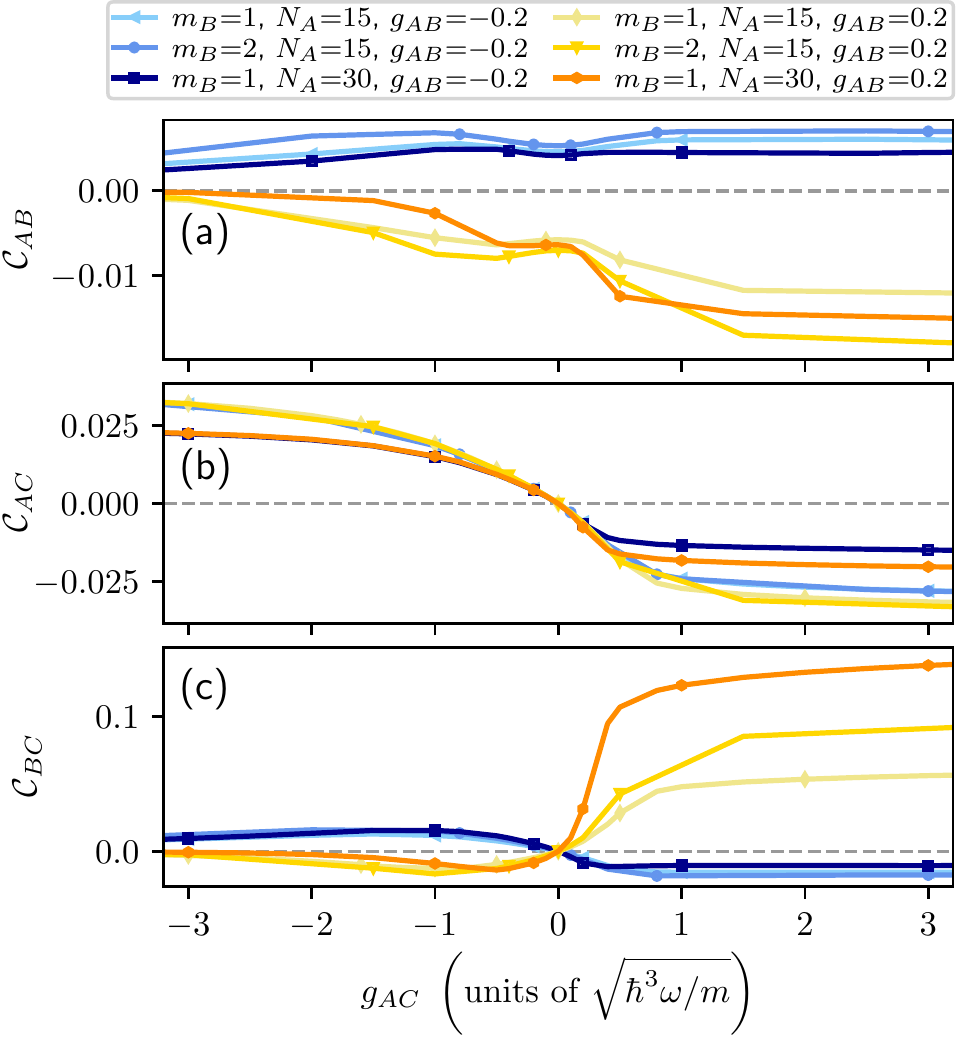}
    \caption{Integrated two-body correlation function [Eq.~(\ref{eq:int_noise_corr})] among (a) the $B$ impurity and the medium, (b) the $C$ impurity and the medium and (c) between the impurities as a function of the intercomponent interaction strength $g_{AC}$. 
    In all panels, we consider fixed $g_{AB}=-0.2, 0.2$ as well different masses of the $B$ impurity (simultaneously setting $\omega_B=\sqrt{m_A/m_B}$) and atom numbers of the medium (see legend), while keeping constant the mean-field interaction $N_Ag_{AA}$. The gray dashed line in panel (c) marks $\Delta \braket {r_{BC}}=0$. }
    \label{fig:mB2_NA30}
\end{figure}

Let us demonstrate the generalization of our results in the main text when the impurities are mass-imbalanced or the bosonic medium contains a larger number of particles. For this purpose, we focus on the behavior of the intercomponent correlations which can be quantified through the integrated correlation function [Eq. (\ref{eq:int_noise_corr})] presented in Figure \ref{fig:mB2_NA30} for different system parameters. 

In general, increasing the mass of an impurity disturbs the cloud of the bosonic gas to a larger degree which should eventually lead to an enhanced impurity-medium correlation. This is indeed evident in Figure~\ref{fig:mB2_NA30}(a) where the integrated correlation function, $\mathcal{C}_{AB}$, is increased as compared to the mass-balanced case, thus testifying an overall larger degree of entanglement. Furthermore, since the correlation between the $C$ impurity and the bath is not affected by the change of $m_B$ [Figure \ref{fig:mB2_NA30}(b)], the larger $\mathcal{C}_{AB}$ leads to a stronger mediated correlation between the impurities, see e.g. $\mathcal{C}_{BC}$ in Figure \ref{fig:mB2_NA30}(c). The latter naturally leads to an amplified impurities' induced interaction for increasing $m_B$. In particular, for $g_{AB}=0.2$ and strong repulsive $g_{AC}$, where $\mathcal{C}_{BC}$ features the largest increase.

Next, we concentrate on the mass-balanced system but consider a larger number of bath particles and in particular $N_A=30$, while maintaining the same mean-field interaction, i.e., $N_Ag_{AA}=\mathrm{const}$. As it can be seen, the impurity-medium correlations, as captured by $\mathcal{C}_{AB}$ and $\mathcal{C}_{AC}$, are reduced compared to the reference case $N_A=15$, $g_{AA}=0.2$ [Figure~\ref{fig:mB2_NA30}(a), (b)]. This is attributed to the smaller intra-species coupling strength $g_{AA}=0.1$ resulting in a decrease of the respective intra-species correlations among the bath particles. However, the mediated correlations among the impurities $B$ and $C$ are clearly enhanced when $g_{AB}$ and $g_{AC}$ are both repulsive, see Figure~\ref{fig:mB2_NA30}(c). In this sense, a larger number of bath particles featuring a decreasing intraspecies interaction is associated to a reduction of intraspecies correlations of the bath and impurity-medium ones but enhances to a certain degree the mediated correlation between the impurities. This behavior hints towards a complicated correlation transfer mechanism to the impurity-impurity subsystem which deserves further future investigations. Nevertheless, a systematic finite size scaling analysis in terms of the atom number in the bath is required in order to deduce the robustness of our findings. However, we expect that the main features of the impurities dressing, e.g. the crossover from a correlated to an anti-correlated behavior (associated to attractive and repulsive induced interactions as discussed in Sections \ref{sec:corr} and \ref{sec:induced_interactions}), and the existence of the impurities bound states for attractive interactions are retained for larger number of bath atoms.

\section{Estimating the importance of correlations on the many-body wave function}
\label{ap:nstate}

\begin{figure}[t]
    \centering
    \includegraphics[width=0.8\linewidth]{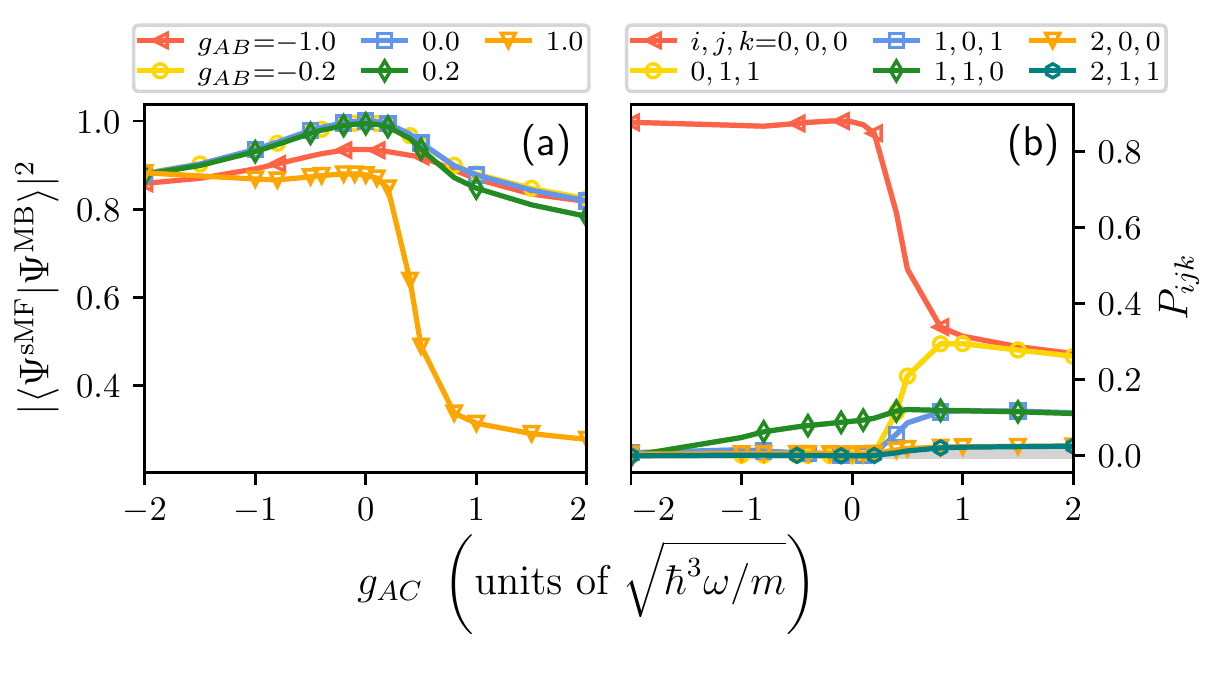}
    \caption{(a) Fidelity between the many-body wave function, $|\Psi^{\mathrm{MB}} \rangle$ (including all emerging intra- and intercomponent correlations) and the species mean-field wave function $|\Psi^{\mathrm{sMF}}\rangle$ where intercomponent correlations are neglected. The reduction of the overlap from unity for finite interactions evinces the participation of intercomponent correlations. (b) Probability amplitude $P_{ijk}$ denoting the overlap of a three-component time-independent basis $|\psi_{i}^A\rangle|\psi_{j}^B\rangle|\psi_{k}^C \rangle$, constructed from the eigenstates of an effective species Hamiltonian (see main text), with the many-body wave function $|\Psi^{\mathrm{MB}} \rangle$. Apparently, energetically higher-lying excited states possess substantial contribution. Probability amplitudes which remain below 0.02 within the interaction range $-2.0 \leq g_{AC} \leq 2.0$ are shown as gray lines. The harmonically trapped three component system consists of two non-interacting but distinguishable impurities immersed in a bosonic gas of $N_A=15$ atoms with $g_{AA}=0.2$.}
\label{fig:nstate}
\end{figure}

To expose the impact of intercomponent correlations at different interaction regimes on the level of the many-body wave function we analyze the fidelity $\left|\langle \Psi^{\mathrm{sMF}}| \Psi^{\mathrm{MB}}\rangle\right|^2$, see Figure~\ref{fig:nstate}(a). Here, $|\Psi^{\mathrm{MB}}\rangle$ denotes the full many-body wave function where all emergent inter- and intracomponent correlations are taken into account, while $|\Psi^{\mathrm{sMF}}\rangle$ refers to the species mean-field wave function which ignores all intercomponent correlations. Naturally, the fidelity is unity when the species are non-interacting, i.e., $g_{AB}=g_{AC}=0$, since in this scenario intercomponent correlations are \textit{a-priori} prohibited. However, the fidelity decays for increasing impurity-medium coupling strengths as intercomponent correlations are triggered in this case. The largest deviation between the many-body and species mean-field wave functions occurs in the parameter region corresponding to the coalescence of the impurities, i.e., for strongly repulsive $g_{AB}$ and $g_{AC}$.

Further understanding of the respective correlation mechanisms can be delivered by identifying the participating microscopic configurations. For this reason we construct the species function eigenbasis $|\psi_{i}^A\rangle|\psi_{j}^B\rangle|\psi_{k}^C \rangle$ obtained by calculating the eigenfunctions of an effective species Hamiltonian [cf. Eq.~(\ref{eq:species_hamilt})] characterized by the effective potential defined in Eq.~(\ref{eq:effpot})\footnote{
The impurities eigenstates are found by solving the corresponding one-body Hamiltonian, while the eigenstates of the effective bath Hamiltonian, consisting of $N_A$ particles, are determined via improved relaxation~\cite{meyer2003}.}.
As basis for the bath we take the ground and the energetically two lowest excited states of the effective potential into account, while for the two impurities we consider the corresponding energetically lowest six eigenstates leading to a total number of 108 three-component basis states $|\psi_{i}^A\rangle|\psi_{j}^B\rangle|\psi_{k}^C \rangle$.

The respective probability amplitudes $P_{ijk} = \left|\left(\langle \psi_{i}^A|\langle\psi_{j}^B|\langle \psi_{k}^C| \right) \Psi^{\mathrm{MB}}\rangle\right|^2$, with $|\Psi^{\mathrm{MB}}\rangle$ being the full many-body wave function, are presented in Figure \ref{fig:nstate}(b) for $g_{AB}=1.0$ and varying $g_{AC}$. Notice that the state $|\psi_{0}^A\rangle|\psi_{0}^B\rangle|\psi_{0}^C \rangle$, denoting the case in which each species occupies the ground state of the effective species Hamiltonian, represents the three-body ground state obtained with a sMF ansatz. Consequently, $P_{000}=\left|\langle \Psi^{\mathrm{sMF}}| \Psi^{\mathrm{MB}}\rangle\right|^2$
(cf. Figures~\ref{fig:nstate}(a) and (b) for $g_{AB}=1.0$). In general, it is observed that finite interactions yield a non-negligible population of energetically higher-lying excited states. Importantly, this behavior becomes enhanced in the coalescence regime, i.e., for strong repulsive $g_{AB}$ and $g_{AC}$. This means that there are several macroscopically occupied basis states reflecting the significant intercompoment entanglement (cf. Figures \ref{fig:corr} and \ref{fig:phase_diag_ent}).

\end{appendix}

\bibliography{literature.bib}

\nolinenumbers

\end{document}